%% file: main.tex
\documentclass[12pt, draftclsnofoot,onecolumn]{IEEEtran}
\usepackage[utf8]{inputenc}
\usepackage{csquotes}
\usepackage{xcolor}
\usepackage[english]{babel}
\usepackage{array}
\usepackage{float}
\usepackage{enumitem}
\newcolumntype{P}[1]{>{\centering\arraybackslash}p{#1}}
\usepackage{multirow}
\usepackage{graphicx}
\usepackage{comment}
\usepackage{subfigure}
\usepackage{caption}
\usepackage{ulem}
\usepackage{url}
\makeatletter

\newcommand{\prk}[1]{{\color{blue}Professor: \textit{ #1}}} 
 \newcommand{\red}[1]{{\color{red} #1}}
\newcommand{\sg}[1]{{\color{red}Santosh: #1}}

\begin{document}

\title{TERRA: Beam Management for Outdoor mm-Wave Networks}

\begin{abstract}

mm-Wave communication systems use narrow directional beams due to the spectrum's characteristic nature: high path and penetration losses. The mobile and the base station primarily employ beams in line of sight (LoS) direction and when needed in non-line of sight direction. Beam management protocol adapts the base station and mobile side beam direction during user mobility and  to sustain link during blockages. To avoid outage in transient pedestrian blockage of the LoS path,  the mobile uses reflected or NLoS path available in indoor environments. Reflected paths can sustain time synchronization and maintain connectivity during temporary blockages. In outdoor environments, such reflections may not be available and prior work relied on dense base station deployment or co-ordinated multi-point access to address outage problem.

Instead of dense and hence cost intensive network deployments, we found experimentally that the mobile can capitalize on ground reflection. We developed TERRA protocol to effectively handle mobile side beam direction during transient blockage events. TERRA avoids outage during pedestrian blockages 84.5 $\%$ of the time in outdoor environments on concrete and gravel surfaces. TERRA also enables the mobile to perform a soft handover to a reserve neighbor base station in the event of a permanent blockage, without requiring any side information unlike the existing works. Evaluations show that TERRA  maintains received signal strength close to the optimal solution while keeping track of the neighbor base station.

 \end{abstract}
  \author{\IEEEauthorblockN{Santosh Ganji, Jaewon Kim, Romil Sonigra and  P. R. Kumar}\\
\IEEEauthorblockA{Texas A\&M  University}}
 
\maketitle
\input{introduction}

\input{challenges}
\input{blockage_measurements}
\input{experiments}

\input{protocol}

\input{evaluation}

\input{results}
\input{related_work}

\bibliographystyle{IEEEtran}
\bibliography{IEEEabrv,references}
\end{document}

%% file: introduction.tex
\section{Introduction}

High path loss in millimeter wave (mm-wave) bands necessitates both the mobile and the base station to communicate using narrow directional line-of-sight (LoS) beams \cite{pathloss}.  
The LoS direction changes as the user move. To remain connected, the base station and the mobile manage beam directions to counter user mobility. Another challenge to mm-wave communication is that the human body significantly attenuates radiation \cite{blockage_attenuation,blockage_rappaport}. Pedestrian blockers cause the narrow directional LoS beam inoperable. mm-wave device must adapt the beam direction to overcome the challenges caused by blockage and user mobility.

When a pedestrian obstructs the mm-wave Line of Sight (LoS) link, the received signal strength drops by about 15 dB \cite{unblock}. Even though such blockage events are temporary and last only about two hundred milliseconds \cite{unblock}, they create link outages as $96.8\%$ of the signal is lost. The signal amplifier cannot improve SNR as both signal and noise are equally amplified. The mobile loses connectivity during such temporary blockages. After the temporary blockage, it has to reconnect like a new user, which in technologies like 5G NewRadio takes about a second \cite{_2017_nr1},
disrupting low latency applications like virtual reality and online gaming \cite{VR_latency}. In an indoor environment, a usable non-line-of-sight (NLoS) path is available \cite{unblock}.

Presuming that reflected NLoS beams are scarce in outdoor environments \cite{beamspy,UDN}, it has been suggested that the solution to avoid disruption is to handover to another base station.  However, handover for every pedestrian blockage requires high deployment density. The study \cite{Ish} suggests 200 base stations per Sq. KM is necessary to meet the latency requirements. Such high deployment density is expensive.

Based on outdoor signal measurement studies under the pedestrian blockage, we show that there is frequently a ground reflection from hard surfaces, and propose a protocol called TERRA. It employs such a ground-reflected NLoS path as a control channel to sustain critical time synchronization with the base station throughout a temporary blockage. This makes it possible to revert to the LoS beam as soon as the blockage concludes. TERRA efficiently maintains the beam identity of such a reserve NLoS path at all times, and refreshes it as needed, so that it is always ready for blockage since blockage can happen at any time. Experiments show that TERRA avoids outage events 84.5$\%$ time during pedestrian blockages.
When either blockage does not disappear or the current base station is out of range, handover to another base station is the only viable choice for the mobile to continue communication. TERRA also addresses the challenges in a handover event that arise from the directional nature of beams. 

 To avoid hard handover where the mobile needs to connect to the next base station as if it is a new user, TERRA must manage mobile beams both with the serving and the neighbor base station. TERRA first needs to determine neighboring base stations for potential handover targets. And for the entire transition process, the protocol must ensure that it has aligned beams at the detected neighbor base station. The mobile must also manage beams with the serving base station till the transfer of connection.

Prior work \cite{twc_beamsurfer,agilelink} corrects the beam misalignment that happens with user mobility. While the mobile can continually adapt its beam during user mobility with the serving base station with cooperation from the serving base station \cite{twc_beamsurfer,agilelink}, the neighbor base station neither adapts its beam nor provides any assistance to the mobile for receive beam adaptation, as the mobile is yet to establish communication with it. Terra uses information only in the radio domain, as is desirable, without requiring any additional sensors. Experiments on TERRA show that it maintains the received signal strength of a neighbor base station within 3 dB of an omniscient oracle. We have also evaluated the performance of TERRA under several pedestrian mobility patterns.

The rest of this paper is organized as follows. We explain the challenges in outdoor beam management in Section \ref{challenges}. In Section \ref{measurements}, we present our outdoor experiments during pedestrian blockages. Section \ref{cell search experiments} shows the search overheard to discover the neighbor base station. We present our protocol in Section \ref{protocol_section}.  Section \ref{evaluation_section} presents TERRA's efficacy in various user mobility scenarios.

%% file: challenges.tex
\section{The challenges of beam management for blockage resiliency
and handover}\label{challenges}
In this section, we elaborate on the challenges in managing beams during pedestrian blockage and in handover. We also provide the outlines of the solution which are further detailed, experimentally verified and validated in the subsequent sections.

\subsection{Pedestrian Blockage}
In the case of omnidirectional transmission, the environment scatters electromagnetic radiation 
% from an omnidirectional transmitter 
in all directions and an omnidirectional receiver can capture 
% this incoming radiation i.e., 
the multi-path components of the transmitted signal. In contrast, due to directional transmission, there are fewer distinct multi-path components in the mm-wave bands. A narrow directional radio receiver beam can only receive signal components that arrive within a small angular spread of a beam direction. A 32x32 element uniform planar array can produce beamwidths as narrow as $4^\circ$. To discover either an LoS or such NLoS paths, the base station sweeps beams within a sector, and the mobile receiver similarly performs a spatial scan.  
% \san{Can we make this 4-6 dB and say that prior works observed greater loss in NLoS components as didn't include ground path}  \prk{Why does the ground reflection have only a 4-6 dB loss (lesser than 10dB loss you just mentioned)?}\san{NLoS are thought of as ambient reflections apart from ground, as reflections from concrete buildings are not studied and significant absorption from normal building walls (not concrete) hence 10 dB for NLoS. In case of hard surface like concrete as radiation cannot penetrate, it gets reflected adhering to some form of energy conservation law.}

% \sout{The receiver may discover an LoS path using wider beams but it cannot discover NLoS paths as they have much lower signal strength.}

% \sout{In the sub 6-GHz band, an obstacle interposed between the base station and the mobile cannot block the transmission
% as the receiver can capture a large number of multipath components.} \sout{However,} 
In the mm-wave band, an interposed pedestrian does obstruct the narrow directional LoS beam.
% due to high oxygen absorption loss. 
The human body  attentuates the signal by 15 dB, resulting poor RSS.
% during blockage events 
With most of the signal energy lost and receiver's amplifier cannot improve SNR. Since the mobile receives weak signal energy and amplifiers cannot improve SNR,  link suffers from outage during pedestrian blockage event.

The mobile is left with one of two choices to continue communication with the network -- to switch to an NLoS path if such a path exists between the base station and mobile, or to perform handover to a neighboring base station (or otherwise employ a neighboring base station through, say, coordinated multipoint transmission). 
% and are not available in open outdoor environments. The mobile is unaware of its ambient environment i.e., whether it is indoor or outdoor without performing an environment scan. 

% The base station and mobile need to adapt their LoS beams to compensate for user mobility. They use phased antenna arrays that electronically steer the direction of radio beams. As the size of the array increases i.e., as the number of antenna elements increases, the directivity gain increases, and the resulting radio beam has a smaller beamwidth. For example, a 32x32 element uniform planar array can produce beamwidths as narrow as $4^\circ$. 
Pedestrian blockage is sudden and unpredictable \cite{twc_beamsurfer}.
To avoid outage by employing NLoS path, 
% it is necessary for the mobile to have a backup NLoS beam direction that it can switch to.
the mobile must therefore always have in hand an NLoS beam direction that it can quickly switch to. 
If the mobile has no backup NLoS path to use in the event of blockage, link outage occurs, and the mobile gets disconnected from the base station.

The disconnected mobile will need to re-perform an initial network access procedure,  as though it were a new user, which takes several seconds, due to the following.
To acquire new users, base stations periodically sweep directional beams with reference signals and 
% {\color{red}\sout{so that a mobile receiving them can IS THIS CORRECT?}} 
broadcast information such as cell and network identity. A mobile also sweeps through all its receive beams to discover at least one of the base station's beams when it is pointed towards it.
The number of receive beams increases with the reciprocal of the
 beamwidth. To complete a bi-directional connection, the mobile transmits a random preamble in the same direction as the discovered base station's beam, and awaits a response. After physical layer procedures to establish reliable data communication, the network authenticates the mobile before granting network access. This complete procedure takes several seconds \cite{Initial_access_delay}.

 On the other hand, were the mobile to have an NLoS path in hand when blockage occurs, it can use that to sustain connectivity in the following way.
The NLoS path has 
% \prk{4dB or \sout{10dB}??} \san{Can we update the number later saying, earlier NLoS paths were thought only high loss however, it is less if we take advantage of NLoS paths from ground?} 
lesser RSS than the LoS path, not enough to sustain the earlier high data rate communication of the LoS path. However, the NLoS path can continue to
sustain time synchronization between the mobile and the base station.
This is critical since it allows the mobile to revert to LoS communication without delay as soon as the blockage disappears, and, typically, such blockage is temporary and only lasts a few hundred milliseconds \cite{unblock}. For such recovery from temporary blockages, it is critical
that it be performed without requiring any out of band communication, and this is exactly what the NLoS path makes possible.

The critical issue is therefore: Do such NLoS paths exist outdoors? \cite{outdoornlos}
Unlike indoor environments, there is not a multiplicity of surfaces when outdoors. One that is always there though is the \textit{ground}.
% It has been experimentally shown that NLoS paths between the base station and the mobile typically do exist in indoor environments
% \cite{unblock,Ish,beamspy,BeamSurfer}.
% \san{According the references and in general literature, NLoS is taken as ambient reflections, not ground reflections, none of them studied ground reflection, only from walls, indoors, as the path loss is less between LoS and NLoS, receive finds wall reflected radiation, where as in outdoor, reflected path from a neighbor building might be far and can be found only if one such path exists}
% \san{I have updated related work, Earlier people observed high loss in NLoS path as they ignored considering ground reflected NLoS path. So now, loss is only 4 to 6 dB both indoors and outdoors. In this particular work we focus on using ground NLoS path for outdoors, so we can avoid dense deployment?}
% \prk{I am confused. Did all these references discover that there are usable indoor reflections? What is the exact contribution of these references? How did they propose using the NLoS beams to overcome blockage? How is it different from your solution?}
% % In indoor environments, \cite{unblock,Ish,beamspy,BeamSurfer} have suggested harvesting NLoS paths to preserve the link between the base station and the mobile during transient blockage events. 
% \prk{In fact you are saying that the outdoor reflection is stronger than indoor NLoS since it is only a 4-6 dB loss compared to 10dB loss indoors. Can you please clarify?} 
Indeed, ground reflections are used to shape glide paths for aircraft instrument landing systems \cite{Glide_Scope}.
% {\color{red} Cite a good reference.}. 
% \sout{Outdoor reflections have also been investigated for communication \cite{NLOSmain1,GRmodel1}.}
There is also a report of ground reflections at mm-wave frequencies in \cite{NLOSmain1} in an outdoor environment.  However, the potential for this to help with mm-wave communications, especially blockage, has 
apparently not been
pursued. 
% \prk{How is your investigation different from these two investiations? How did they propose to use the reflected path?} 
Motivated by this possibility, we have conducted a measurement campaign in the 60GHz band to determine whether mm-wave signals are reflected from the ground, 
and whether they are usable during blockage events.

\begin{figure}
    \centering
    \includegraphics[width=.5\linewidth]{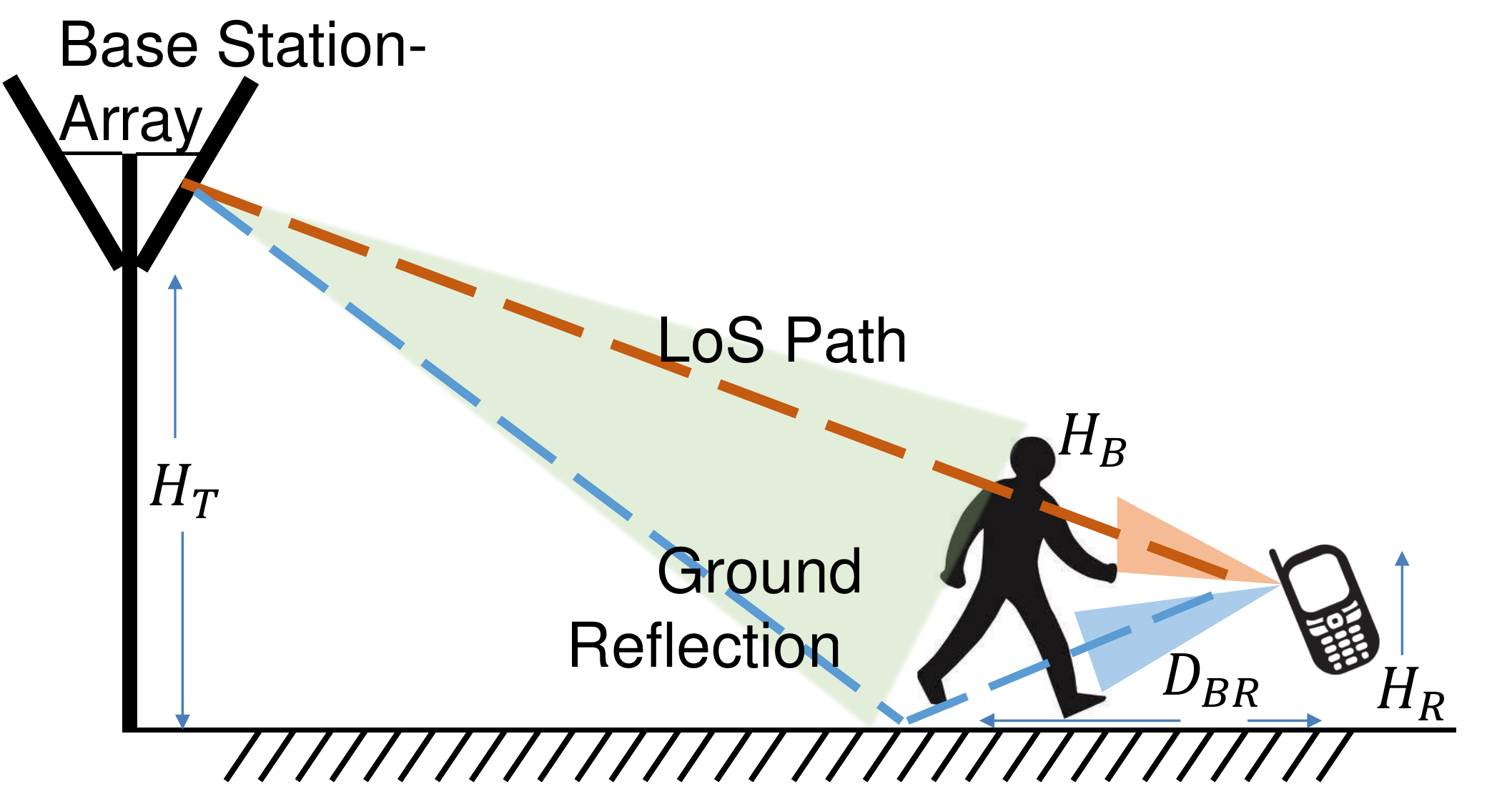}
    \caption{Ground Reflection}
    \label{GR}
\end{figure}

In Section \ref{measurements}, we report on the results of the measurement study
that show that mm-wave signals are  reflected from outdoor surfaces such as concrete and gravel with a loss of just 4-6 dB over LoS. 
Also important for base station to hand-held mobile communications is that base stations are usually deployed with a slight downward tilt, as shown in Fig. \ref{GR}, and are equipped with phased arrays that steer beams. The hand-held mobile's receiver can therefore capture these ground reflections.  
We also report in Section \ref{measurements} on
the link measurements with human blocking of the LoS link between the base station and the mobile.  
% {\color{red} Do you want to stick to outdoors in this paper?, and ceramic tiles.} We have also observed strong ground reflections in both indoor and outdoor environments. 

% Also, very important for our purposes is that base stations are usually deployed with a slight downward tilt, as shown in Fig. \ref{GR}, and are equipped with phased arrays that steer beams, the mobile's receiver can therefore capture these ground reflections, which are moreover in the same azimuth direction as the LoS path. 
% % \prk{But we cannot reduce the search process and time for an NLoS beam since the mobile phone may be in a tilted position and it does not know what is azimuth and what is elevation}. \san{agreed} 
% To investigate if such reflections are usable during blockage events, we performed link measurements with human blocking of the LoS link between the base station and the mobile.  
% % \prk{including ?? Mention the surfaces tested. If you found out that some specific surfaces have even lower loss, could you mention the specific losses of different surfaces in the Tables for Concrete Surface and Gravel in the next section possibly}. 
% % \prk{``but this is not reliable" Is this OK to say?}. 
% Our experiments and observations are detailed in Section \ref{measurements}. 

In cellular communication systems, the base station schedules data transmission and reception opportunities. 
Time synchronization helps mobile adjust its timeline to that of  base station's. 
To align the mobile's time line, the base station transmits synchronization signals. Through signal correlation, the mobile determines the temporal location of these signals in the captured over the air samples and adjusts its timeline. A good received signal strength and high enough
SNR are necessary to improve the accuracy of signal processing algorithms and help achieve tight timing alignment with the base station. 
 Any timing offset induces packet errors in both uplink and downlink \cite{ofdm_time_sync_issues} as the mobile's transmissions fall out of the base station's listening window and vice versa. The
 mobile must therefore continuously adapt its timeline to account for the propagation delay with user mobility. It is for this reason that TERRA switches to a NLoS beam to maintain time synchronization throughput the temporary
 blockage period. This allows it to revert to LoS communication as soon as the blockage ends.

\subsection{Beam Management for handover}
\subsubsection{Brief overview of 5G mm-Wave handover process}
When a mobile user moves to the boundary of the currently connected base station's coverage region, called a ``cell", the mobile experiences degraded radio conditions. At the edge of the cell, the received signal strength is weak and hence Signal to Noise Ratio (SNR) is poor. When SNR is bad, packet decoding fails. Cellular technologies like 5G {can} use hybrid packet retransmissions to use previously transmitted bits to decode a message before sending the original message again with a reduced code rate. Either way, when a packet is decoded in error, the recovery mechanisms increase communication latency. A similar situation occurs when a LoS beam is permanently blocked by a building.

Under such edge radio or permanent blockage conditions, improve link performance, the mobile searches for a neighbor base station. While omni-directional mobile receivers need to perform only a \textit{frequency scan} to discover the neighbor base station and initiate the handoff process, in the mm-wave bands, the mobile uses narrow directional beams and therefore needs to perform a \textit{spatial scan} to discover a  neighbor base station. 
% Later,  mobile decodes the broadcast information that contains necessary information to initiate connection transfer.
 A 5G mm-wave base station periodically sweeps broadcast information using narrow directional beams \cite{_2017_nr1}.  Broadcast messages help mobile discover the neighbor base station's timeline. The mobile must adjust its timeline to align with the neighbor base station. This process called ``time synchronization" is the foremost step a mobile performs before starting the initial access protocol. During initial access procedures, the mobile informs the neighbor base station of its presence in the coverage area. The mobile transmits an uplink preamble signal in a listening window of the base station and anticipates a response. Tight time synchronization at the mobile ensures that the sent preamble reaches the base station. 
  
  Serving base station cannot help a mobile with the time schedules of a neighbor base station. Without strict time synchronization among the base stations in the network, a base station cannot have the knowledge of timing of another. For example, serving base station precisely time synchronize with neighbor base station to convey to its mobile when a beam is available from the neighbor base station to initiate communication.

The broadcast messages also carry schedules of when the base station uses a particular beam to listen to a mobile's transmissions. To complete handover, the mobile must transmit precisely at the instants when a neighbor base station is listening in the direction of the beam discovered after the spatial scan. Moreover, cellular standards require the mobile to choose time-frequency resources for preamble signals randomly. The neighbor base station listens to the sent preamble and responds when there is no resource collision. 
A response from the base station to a preamble is necessary for the mobile to advance further in the handover protocol. The mobile waits for a response for a pre-configured interval, after which it retransmits a new preamble \cite{Initial_access_delay}
%Unlike sub-6-GHz Omni-directional cellular systems, the mobile must perform the handover using directional radio beams. 
Upon receiving a response to the preamble signal, the neighbor base station and mobile exchange critical control plane messages for user authentication and connection transfer. The mobile must maintain a highly aligned beam throughout this process to avoid handover failure. 

In short, the mobile must search for a neighbor base station, time synchronize and perform initial access procedure to handover. 
Below we elaborate on the main stages in a handover protocol and implications of beam management on those stages.

% To summarize, the mobile performs following steps during a handover event.
% \begin{itemize}
%     \item Search for neighbor base stations.
%     \item Identify a reliable neighbor base station beam.
%     \item Time Synchronize to the neighbor base station.
%     \item Adapt the mobile side beam to continue communicating with base station. 
%     \item Complete initial access procedure and establish radio level connection with the neighbor base station.
% \end{itemize}

%We elaborate the challenges in each of the above mentioned stages.

\subsubsection{Neighbor Base Station Search} \label{subsection_challenges_cell_search}

As mentioned earlier, the serving base station schedules persistent measurement occasions for the mobile to discover neighbor base stations \cite{_2017_nr3}.
 In the granted opportunities, mm-wave mobile performs directional search on frequencies that are communicated by the serving base station.
Mobile first measures signal strength temporarily tuning the radio receiver to  carrier frequencies of neighbor base station and attempts decoding the broadcast information that contains the network related information. 

The mobile searches for a neighbor base station using one receive beam at a time. As   transmit beam schedules of neighbor base station are unknown, mobile uses the same receive beam for the entirety of one beam sweeping interval. In 5G mm-wave network, mobile holds each of its receive beams for 20 ms, the duration in which base station sweeps all its beams once. Mobile has complete freedom on the beamwidth of beams.  Search concludes after discovering a base station and mobile must report signal measurements to the serving base station. Time to discover a base station is impacted with beamwidth, number of beams, and user mobility pattern.

Based on the reported measurements, serving base station then makes final decision on handover. However, such a mechanism does not have any particular advantage. Irrespective of whether mobile or serving base station that decides on the switch, there is no way to evade measurements. Until the switching decision is made, the mobile's receiver needs to toggle between serving and neighbor base stations. Mobile must keep track of the found neighbor base station beam until the handover is initiated. Also, the mobile must keep track of serving base station beam. Failure to track the serving base station beam results in hard handover whereas losing track of neighbor base station beam requires mobile to search again.

\subsubsection{Beam Tracking} \label{subsection_challenges_beam_tracking}

 Mere one-time discovery of a neighbor base station beam is insufficient to complete the handover. The mobile must maintain alignment with the found base station beam to overcome mobility impairments and maintain good received signal strength throughout the transition process. 
This adaptation step, called Beam Tracking, involves the mobile switching its receive beams to maintain high received signal strength. Beam tracking is essential to complete all the handover protocol message exchanges and avoid a hard handover.

As the mobile is yet to establish connection with the neighbor base station before handover, at the time of handover neither does the neighbor base station adapt its beams to counter user mobility nor does it assist the mobile to adapt its beams to preserve beam alignment. The mobile can only rely on its own beam adaptation.
Since this adaptation of its receive beam to the neighbor base station is done
without any communication with the neighbor base station, we call it ``silent tracking."  Silent Tracking is different from adaptation that is done with the serving base station, or which will be done
with the neighbor base station after a connection handover since both of the latter exploit two-way communication with the respective base stations. In connected state, the base station also adapts its beams to counter user mobility and aids the mobile in adapting its receive beams.  This can done with mobile's adaptation procedure in the companion
Beamsurfer protocol \cite{twc_beamsurfer} where the mobile shifts to one of its better adjacent beams when the
RSS of the current beam drops by 3dB. The details can be found in \cite{twc_beamsurfer}.

\begin{comment}

The angle of arrival estimates require mobile to i.e., mobile orientation while making the angle of arrival estimates and after choosing a beam based on the estimates must remain within the beamwidth of chosen beam. However, a phased array with 64 elements forms beams as narrow as $1^\circ$. Given user's hand and body movements, such a requirement is hard to met. 
 Fig. \ref{figure1both}b, shows how a mobile's orientation impacts angle of arrival estimates. At first, it might appear AoA measurements help to align mobile beam but such stringent requirements voids AoA for beam alignments.  As evident from Fig. \ref{figure1both}b., the mobile AoA estimates changes with orientation, adding to that angle measurements are mostly inaccurate in the cell edge radio conditions, makes AoA unreliable for beam management at the cell edge. 
\end{comment}

%\ctxt{This section presents why beam tracking is necessary for longer duration.}

\subsubsection{Time Synchronization, Random Access and Connection Transfer}\label{subsection_challenges_random_access}

 After beam discovery and timing synchronization, the mobile transmits a preamble signal to announce its presence to the {neighbor} base station. 
 The preamble  and the time-frequency resources to transmit the signal are chosen randomly from a set known to the base station.
This step in the initial access procedure is called random access.
 The base station listens {for} all possible preamble resources.  
After listening to the preamble signal, the base station responds and allocates resources for the mobile to complete the rest of the initial access procedure.
The preamble must arrive within the base station's listening window. As the communication at mm-wave bands is directional, the base station listens in a particular direction in each window. The mobile must therefore maintain tight time synchronization with the {neighbor} base station. Upon receiving a response to the preamble signal, both the mobile and the base station exchange several protocol messages to complete the transition. So, the mobile must still maintain a receive beam adapted to user mobility to continue the handover procedures. 
The neighbor base station neither adapts its beams nor provides any assistance to mobile in adapting beams during the initial connection.

 \begin{figure*}[h]
  \centering
 \includegraphics[width=.8\linewidth]{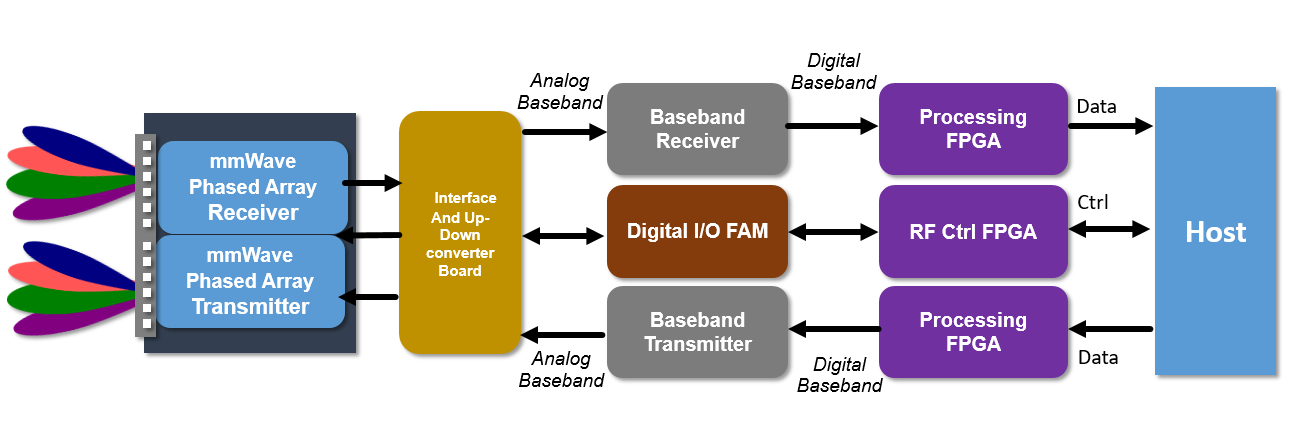}
  \caption{60 GHz Transceiver System }
  \label{figure3}
  \end{figure*}

%% file: blockage_measurements.tex
\section{Measurements Under Pedestrian Blockage} \label{measurements}

We performed signal strength measurements under pedestrian blockage in outdoor built environments with commonly found ground surfaces. The goal of the experiments is to study signal on LoS path during blockage and to explore NLoS paths, in particular ground reflections. NLoS paths can sustain the link and help mobile maintain time synchronization when the LoS path is blocked. 

\begin{figure}
	\centering
	\begin{minipage}{.45\columnwidth}
		\centering
		\includegraphics[width=.9\textwidth]{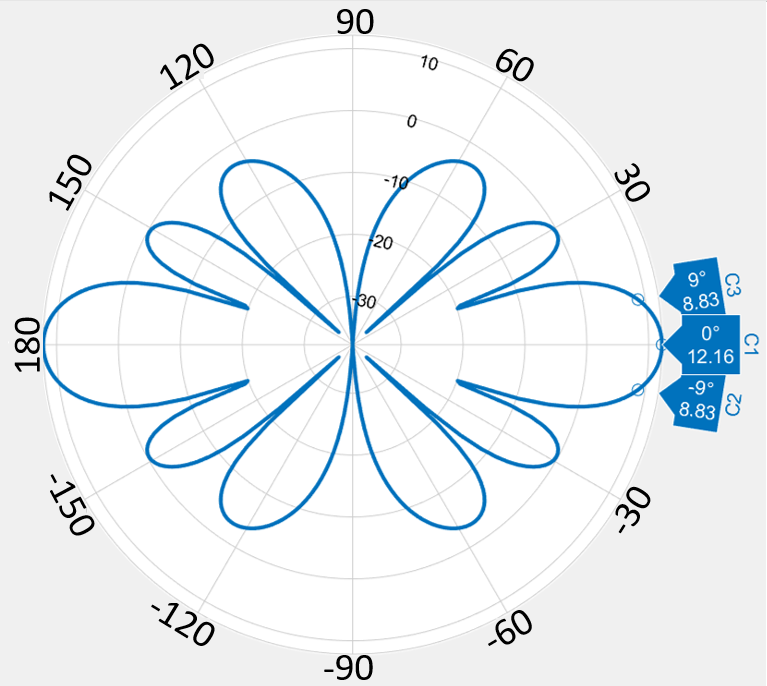}
		\caption{Azimuth Cut}\label{Azimuth}
		%\label{label1}
	\end{minipage}%
	\hfill
	\begin{minipage}{.45\columnwidth}
		\centering
		\includegraphics[width=.9\textwidth]{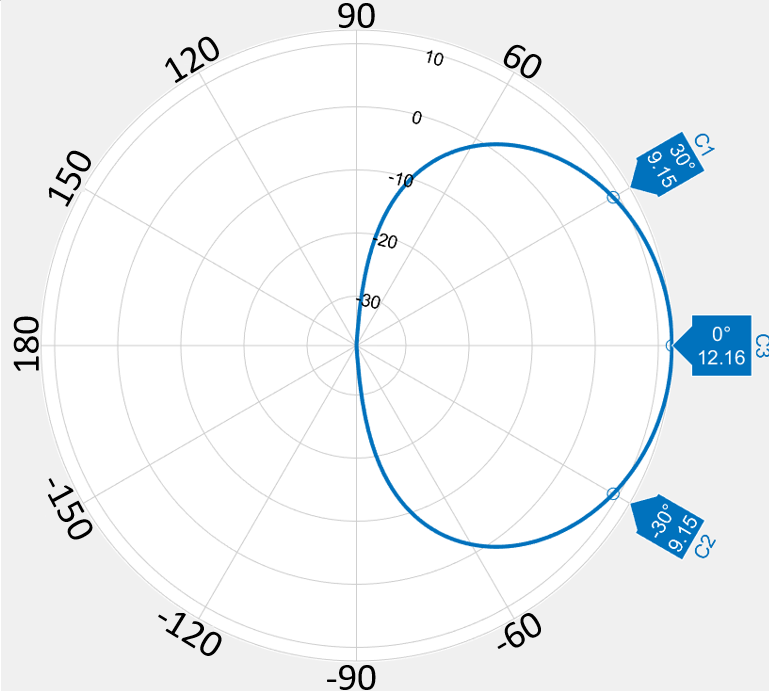}
		\caption{Elevation Cut}\label{Elevation}
	\end{minipage}
\end{figure}

We conducted experiments using  National Instruments 60 GHz software-defined radios. Functional diagram of transceiver is shown in Fig. \ref{figure3}. Baseband IQ sample generation at transmitter and signal processing at the receiver are implemented in FPGA. An analog baseband signal of 2 GHz bandwidth is upconverted to 60 GHz carrier frequency. A 12 element phased array is used both at the transmitter and receiver.
The phase weights for desired radiation patterns are calculated and stored as beam codebooks.  
% {\color{red} Sanrtosh: Please change ``code books"/``code book" to ``codebooks"/``codebook" throughput the paper.} 
 The beam codebook has 25 beams, with narrow beams 
of {azimuth} width approximately 
% {\color{red} Should this be chaged to $18^\circ$, since that is what os specifically mentioned
% below?}
$18^\circ$,
within a $120^\circ$ azimuth sector.  The zenith beamwidth is around $60^\circ$.
Figs. \ref{Azimuth} and \ref{Elevation} present  azimuth and elevation radiation patterns of the bore sight beam. Transmit power is fixed at 20 dBm. The directivity gain of the phased array is 17 dB.
Further details on the transceiver design and implementation are available in \cite{unblock,saha_2017_a}. 

On each surface under study, the transmitter array  was positioned at a height of ${H_T =}2.5$ m above ground level, with the receiver antenna array held about ${H_R=} 1$ m above the the surface. The transmitter and receiver arrays were positioned facing 
% {\color{red} But in practice the mobile may be rotated and not facing the transmitter. How do you know your conclusions also extend to such positions?}
each other, and are placed ${D_{TR}} =6$ m apart. For each scenario, experiments were repeated for two different cases where the transmitter  antenna is tilted towards the ground by $10^\circ$ or $20^\circ$.  This geometry corresponds to potential outdoor deployments where base stations are located higher than mobiles. This tilt is responsible for creating additional reflected directions towards the receiver. Moreover, most of the zenith  beamwidth is directed towards the receiver.

% {\color{cyan} \emph{Mobile Rotation}mm-wave mobile devices have antennas on 3 out of 4 edges \cite{Qualcomm} This might solve rotation issue. When the mobile is not facing, either it is communicating using NLoS or it might have to fallback to sub-6GHz technologies to get desired data rates, as mm-wave link can't meet its data demands}

While the transmitter beam is in the LoS direction of the receiver, the  {RSS} {at the receiver, denoted by $RSS_{LoS}$,} is measured using a {receive} beam that is highly aligned with the transmitter beam. 
% We denote by $RSS_{LoS}$ the signal strength in the LoS direction at the receiver. 
It serves as a reference to calculate total loss suffered by ground reflection. $RSS_{LoS}$ in the experiments was measured to be $-60$ dBm. When a human {body} obstructs the LoS direction by standing in between transmitter and receiver, the RSS drops to $-78$ dBm, which is the noise floor of the receiver, showing that pedestrian obstruction leads to LoS link loss. 
Although the pedestrian blockage is transient, an undesirable outage event occurs at the receiver if communication, and thereby time-synchronization, is not maintained as described in Section \ref{challenges}.

% Let $H_T$ and $H_R$ be the \sout{distances from} \red{heights above} ground level of the transmitter array and receiver array, respectively, $D_{TR}$ the distance between transmitter and receiver, and
Let $H_{B}$ be the height of a pedestrian human blocker. The pedestrian blocker can  obstruct the transmission only when she is close to the receiver. Using ray tracing,  the maximum distance $D_{BR_{max}}$  between blocker and the receiver to obstruct LoS transmissions 
can be calculated as follows:
\begin{equation} \label{eqn1}
	D_{BR_{max}} = D_{TR}*\frac{H_{B}-H_R}{H_T-H_R} .
	\end{equation}
For $H_R=1$m, $H_T=2.5$m,   $H_B$= 1.78m, and $D_{TR}=6$m, $D_{BR_{max}}$ was found to be 3.12m in our experiments.
% \prk{How does the RSS fall of as the distance increases? Is it true that this protocol cna handle self-bockage by the same pedestrian using the mobile? If so, you could say that the distance on that case is much less. What is the typical use case?} \san{I tried self blockage, but it not strictly possible to overcome. Our use case is only pedestrian blockage other than user.}

\begin{table}[ht]
\small 
 \caption{Concrete Surface}
   \label{Concrete}
\begin{center}
\begin{tabular}{ | c | c | c| c|}
\hline
\hline
 Transmitter Tilt  & $RSS_{GR}$ (dBm) &  $D_{BR}$ (m) \\ \hline \hline
$0^\circ$ & -66   & 2\\
$0^\circ$ & -66   & 3\\  
\hline
$10^\circ$ & -64.7   & 2\\  
$10^\circ$ & -64.5  & 3\\  
\hline
$20^\circ$ & -64.1   & 2\\ 
$20^\circ$ & -64  & 3\\ 
 \hline
\end{tabular}
\end{center}

\end{table}

\begin{table}[h]
\small 
 \caption{Gravel Surface}
   \label{Gravel}
\begin{center}
\begin{tabular}{ | c | c | c| c|}
\hline
\hline
 Transmitter Tilt  & $RSS_{GR}$ (dBm) & $D_{BR}$ (m) \\ \hline \hline
$0^\circ$ & -66.1   & 2\\
$0^\circ$ & -65.9   & 3\\  
\hline
$10^\circ$ & -64.8  & 2\\  
$10^\circ$ & -64.4  &  3\\  
\hline
$20^\circ$ & -64.4   & 2\\ 
$20^\circ$ & -64.3  & 3\\ 
 \hline
\end{tabular}
\end{center}

\end{table}

Tables \ref{Concrete} and \ref{Gravel} present the RSS for outdoor reflections from  Concrete and Gravel pathways.

The extra loss incurred by the ground reflection is between 4 and 6 dB.

When both transmitter and receiver phased arrays are parallel to the ground surface, the only ground reflection available to the receiver is from radiation in one-half of the $60^\circ$ elevation beamwidth of the transmitter beam. 
To capture the reflection in this direction, the receiver needs to tilt 
% {\color{red} But in practice the user may position it in any random way.So are the results generalizable?} 
its beams towards the ground while maintaining LoS in azimuth. 
{Indeed we measured slightly more RSS when the receiver is so tilted.}

When the transmitter array is also tilted towards ground, directions with stronger incident radiation get reflected, resulting in higher RSS. The highest $RSS_{GR}$ observed on all three surfaces under study is around -64 dBm. This implies that ground reflected radiation is just 4 dB less than LoS. 
% It is also important to note that ground reflected radiation takes a shorter path to reach the receiver than NLoS paths from the environment. 

Based on the experiments, the following are our main observations:
\begin{itemize}
    \item Pedestrian blockers can create mm-wave link outage, however, NLoS paths to preserve link are available in outdoors too.
    \item Strong ground reflections are available on gravel and concrete built
    % \sout{in both indoor and} 
    outdoor surfaces.
    \item Ground reflections are available in the same azimuth LoS direction at the receiver.
      \item Tilting the transmitter towards the ground helps the receiver with even stronger ground reflections. 
      \item Finally,  when the ground reflected directed path is known to the mobile, there is no need to handover to a neighboring base station in outdoor environments during transient blockage events.
      
\end{itemize}

We note as an aside that after discovering ground reflections outdoors, we also experimented with surfaces indoors, and discovered the presence of indoor ground reflections also when the floor is hard.
Table \ref{Indoor} presents $RSS_{GR}$ averaged over 100 measurements from an indoor surface with concrete tiles. In fact, compared to the indoor NLoS paths reported in  \cite{twc_beamsurfer}, the $RSS_{GR}$ is at least 6 dB higher than RSS of NLoS paths. Therefore it appears that even indoors one can preferably use ground reflections when the floor is a hard surface. TERRA protocol harvests these ground reflected directions, stores in the memory and employs that NLoS path in a blockage event.

\begin{table}[H]
\small 
 \caption{Indoor Floor, Concrete Tiles }
   \label{Indoor}
\begin{center}
\begin{tabular}{ | c | c | c| c|}
\hline
\hline
 Transmitter Tilt  & $RSS_{GR}$ (dBm)  & $D_{BR}$ (m) \\ \hline \hline
$0^\circ$ & -65.7  & 2\\
$0^\circ$ & -66   & 3\\  
\hline
$10^\circ$ & -64.5  & 2\\  
$10^\circ$ & -64.45  & 3\\
\hline
$20^\circ$ & -64.4   & 2\\ 
$20^\circ$ & -64.3   & 3\\ 
 \hline
\end{tabular}
\end{center}

\end{table}

%% file: experiments.tex
\section{Mobility Experiments on Overhead of Beam Search}\label{section_mobility_experiments}
%We performed experiments using a 60 GHz testbed \cite{instruments_2020_introduction}.
{We now address the overhead of TERRA to maintain
a potential neighbor base station to use in case handover is needed. We present the experimental results on the} number of beam searches necessary to discover neighbor base station beams while a user is walking.  We show that search overhead varies with user's location,  orientation of the antenna array, and user mobility pattern.  

 During the pedestrian mobility experiments, the user moves with the phased array in hand. Additionally, to showcase how rotational motion impacts search, a phased array is rotated with angular velocities  {corresponding to} the natural movement of a user's hand. We also present the results in the scenario where two {contending} neighbor base stations are visible to the mobile.

\subsection{Experiment Setup:}\label{subsection_mobility_experiments_experiment_setup}
We use National Instruments' software-defined mm-wave transceivers operating at $60$ GHz for the experiments. One of the transceiver operates as base station, another as mobile. 
%The radio transceiver has several FPGA cards with dedicated baseband processing functions, digital to analog and analog to digital converters as shown in Fig. \ref{figure3}.  These FPGA cards form the data processing pipelines using high-speed interconnects.

%Each transceiver has a $12$ element phased array, $2$ bit phase weights applied to each element. Applying desired phased weights to each, we create beam patterns with varied directional gains and beamwidths. We use a codebook of weights that forms a set of 25 beams spanning $120^\circ$ sector.  

%The uplink and downlink transmissions are time division duplexed. 
%Each slot carries either a reference signal or uplink or downlink traffic. All slot are assigned a traffic flow to have bi-directional.
   
\begin{figure}[h]
    
     \centering
      \includegraphics[width=.40\linewidth]{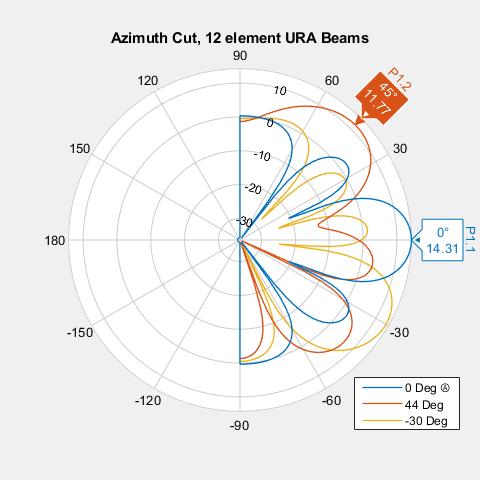}
    
     \caption{Beam patterns of a few  beams from codebook used in the testbed}\label{beampatterns}
\end{figure}
\subsubsection{Data Collection:} \label{data collection}
The base station and mobile use $25$ different beams, each of
beam width approximately $15^\circ$.
These 25 beams cover $-50^{\circ}$ to $60^{\circ}$ azimuth around the boresight of the array.  
% Each beam has a beam width of approximately $15^\circ$. 
On similar lines as existing $5$G cellular standards \cite{_2017_nr1}, the base station transmits a reference signal known to the mobile in $25$ different beams every $20$ milli-seconds. 
Base stations transmits reference symbols in time slots, changing transmitter beams every $8$ slots or $800$ micro-seconds.

The mobile attempts to discover at least one base station beam. 
In doing so, the mobile performs an exhaustive search, switching the receive beams.  The base station's broadcast beam schedules are unknown to the mobile during the beam discovery phase, so the mobile holds a receive beam for $20$ milli-seconds before switching to another. The duration to sweep all the $25$ receive beams at the mobile receiver is $500$ milli-seconds. For beam discovery experiments, each trial lasts for 25 seconds during which the mobile repeats the search. Fig. \ref{beampatterns} presents the azimuth cut of a few beams used in the experiment.
 
 % cell in the experiments is 10 m from the base station.
 
%\red{Is the first sentence needed? I am undecided between yes and no. It may be useful to the reviewer in terms of appreciating the details of the experimentation.} 

The mobile's baseband signal processor correlates the received samples with the reference signal and calculates the received signal strength if the reference signal is detected. We make a note of the number of searches needed to discover a base station beam and the received signal strength of found beam.  
\subsection{Beam Search} \label{cell search experiments}
As part of search process, the mobile measures the received signal strength on each of its receive beams to discover a base station beam. {The search terminates when} one of the receive beams has sufficient received signal strength to decode broadcast information and discovers a neighbor. This beam direction is recorded for potential future handover, and the search 
 is commenced anew.

Sensors providing tilt or pose information of a mobile can help with searches.
They can be especially helpful in the case of searching in a purely zenith direction for a ground reflected beam. However, during initial search or base station discovery, additional side information from sensors gives no advantage.
Also, the particular challenges of utilizing sensors such as
gyroscopes, tilt sensors, and accelerometers,
to obtain angle information like pose have been investigated in 
\cite{_2017_nr1,android_developers,odenwald}.

% pose prediction is complex. IMUs can't help. add papers?

\subsubsection{Beam Search During Walk}
  We conducted mobility experiments to  {determine the} search duration necessary to find a base station beam during pedestrian mobility. We present the following metrics for each mobility pattern; the number of searches required and the received signal strength. For  {the}  human walk experiments, we consider two mobility scenarios. In the first, a user with a phased array in her hand follows a linear trajectory walking at $1 m/s$.  The linear trajectory captures the
  translational motion component of a user walking along a pavement. On the other hand, the free walk has both translational and rotational components, for example when playing games with virtual reality gear.
%   Linear trajectory depicts a user walking on a well-laid footpath or pavement and capture translational motion component of the human walk, whereas free walk has both translational and rotational components like a human walking in a park or playing games with a virtual reality gear. 
The length of each trajectory is $2$ m. We conducted $50$ trials for each scenario, and at five different locations.
  
 To perform a spatial scan with all the 25 beams holding each beam for 20 ms takes half a second. At four different locations, that are {each  half a second long along each trajectory}, we present the variation in the number of searches to discover a base station beam and received signal strength. We number these locations as positions 1 to 4. 
 
A mm-wave base station has only a small coverage region, i.e., a small ``cell". The mobile can listen to transmissions of a base station when the mobile is in the coverage region, and the mobile's receive beam aligns with the base station beam.

\textbf{Linear Trajectory:}  In this mobility experiment, the boresight of the mobile array faces the base station, so that the neighbor base station array and mobile array are facing each other. This experiment shows how the location of the mobile alone impacts the search process. The mobile does not know when the neighbor base station directs a beam towards it. Even though there is only one mobile beam that is in LoS direction at a particular location with the neighbor base station,  number of beams to search to discover LoS vary with time. 

\begin{figure}[t]
 \begin{subfigure}[]
      \centering
         \includegraphics[width=.45\linewidth]{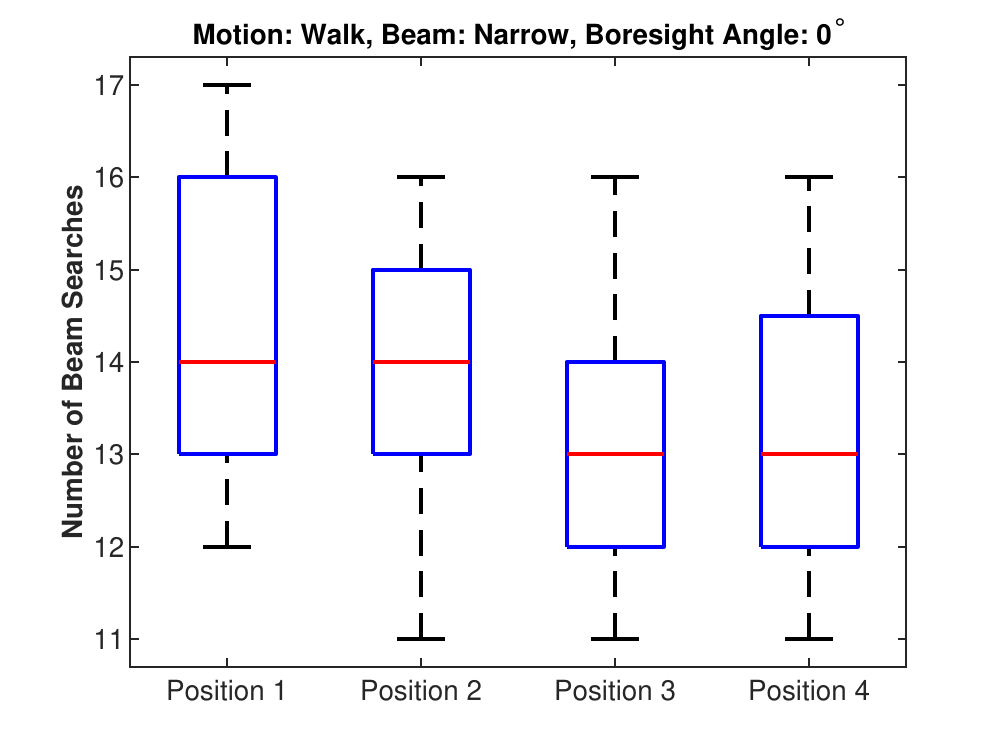}
    \end{subfigure}
           \hfill
     \begin{subfigure}[]
     \centering
      \includegraphics[width=.45\linewidth]{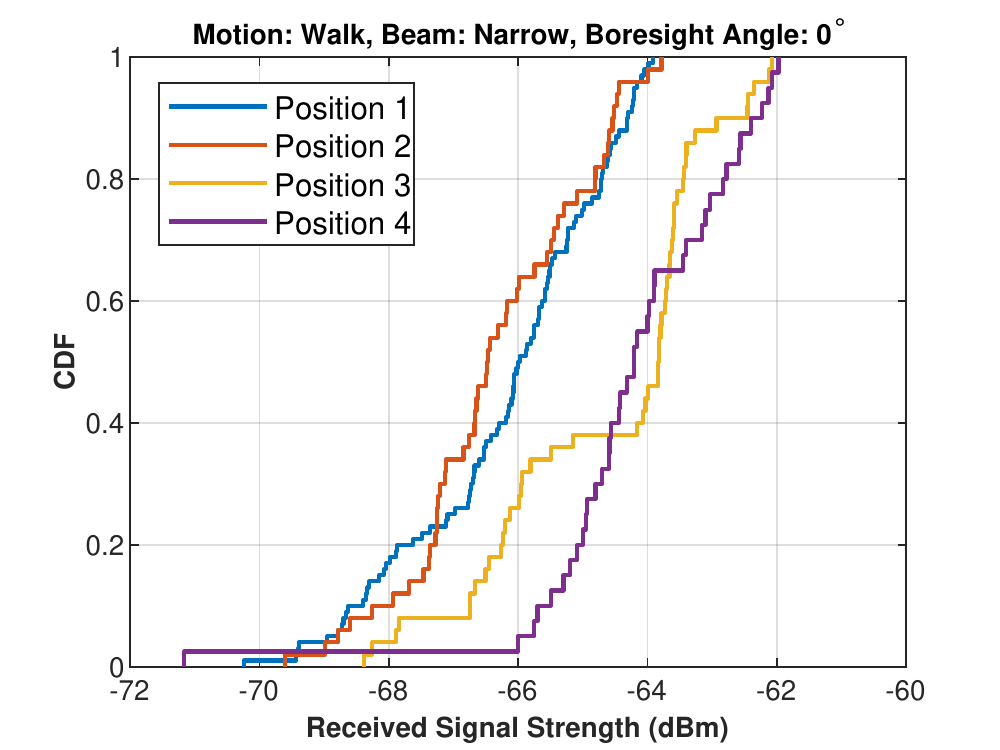}
     \end{subfigure}

     \caption{Search during linear translational motion}
         \label{lt_boresight_0}
     \end{figure}

Along the trajectory, the number of searches necessary to detect the line of sight base station beam varies with position. In Fig. \ref{lt_boresight_0}a, the median number of beam searches is 14 at positions 1 and 2, and 13 at positions 3 and 4. As the perfect beam alignment may not occur all the time and due to slight difference in gains of beams in our code book, signal strength in our experiments is not uniform across all trials. The median received signal strength observed from Fig. \ref{lt_boresight_0}b at positions 1 and 2 is 2 dB less than that of positions 3 and 4. We find it reasonable to expect variation in signal strength due to mobility in directional mm-wave communication systems.

     \begin{figure}

     \begin{subfigure}[]
      \centering
         \includegraphics[width=.45\linewidth]{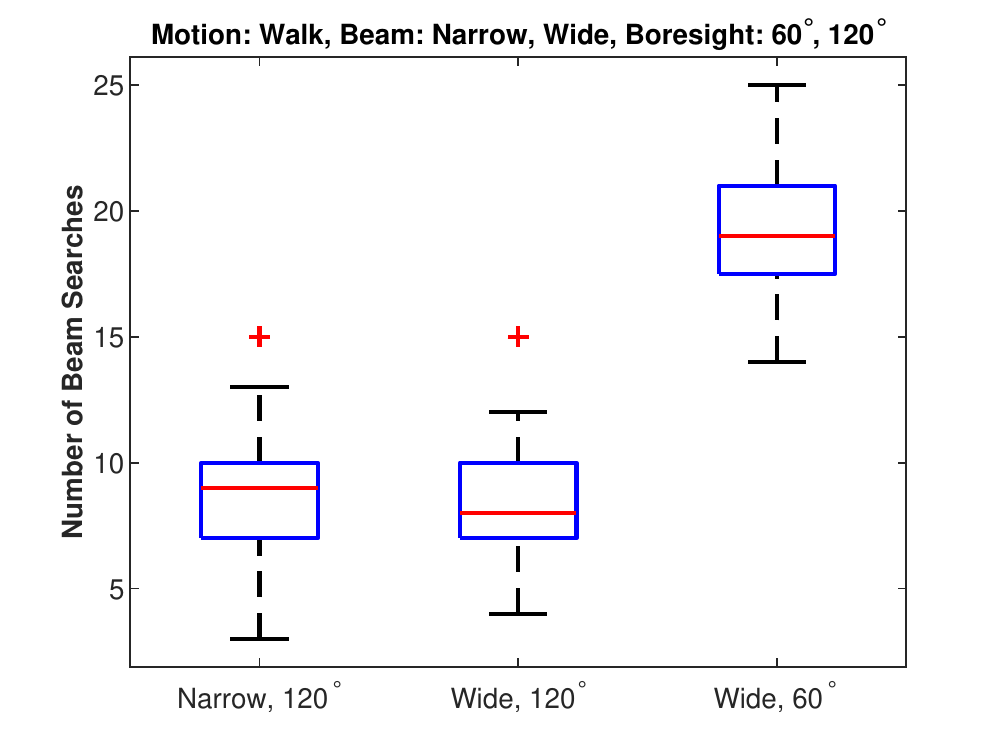}
           \end{subfigure}
         \hfill  
     \begin{subfigure}[]
     \centering
      \includegraphics[width=.45\linewidth]{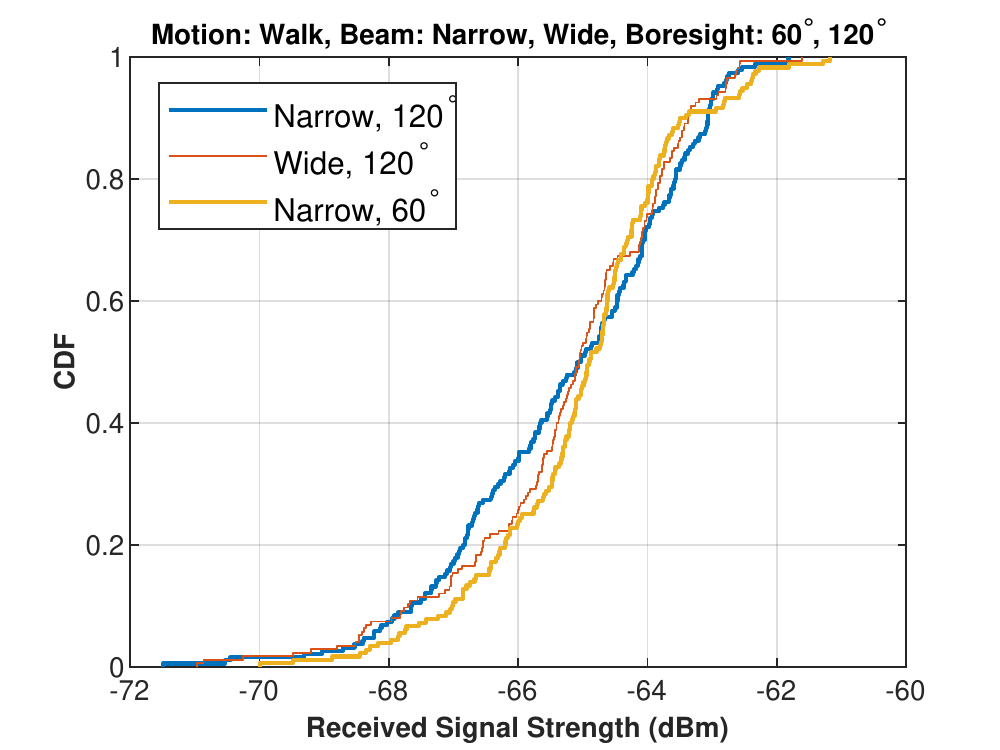}
     \end{subfigure}

     \caption{Impact of array boresight direction on search}
     \label{lt_boresight_60_120}
  \end{figure}

\textit{Impact of Boresight Direction:}
The direction of the antenna array i.e., the array boresight, also has an impact on the spatial scan. As shown in Fig. \ref{lt_boresight_60_120}, when we repeat the experiments with different boresight directions of the phased array, we observe the median number of beams to search when the array points towards $60^\circ$ azimuth is 20, while it is 9 for $120^\circ$.  The median received signal strengths in Fig. \ref{lt_boresight_60_120}b are slightly lower compared to Fig. \ref{lt_boresight_0}b, the case where both the base station and mobile are facing each other. The reason is the irregularities in the gains across beams in the codebook.  The codebook needs careful design to have uniform gains.

\begin{figure}[H]
    \centering
    \includegraphics[width=.45\linewidth]{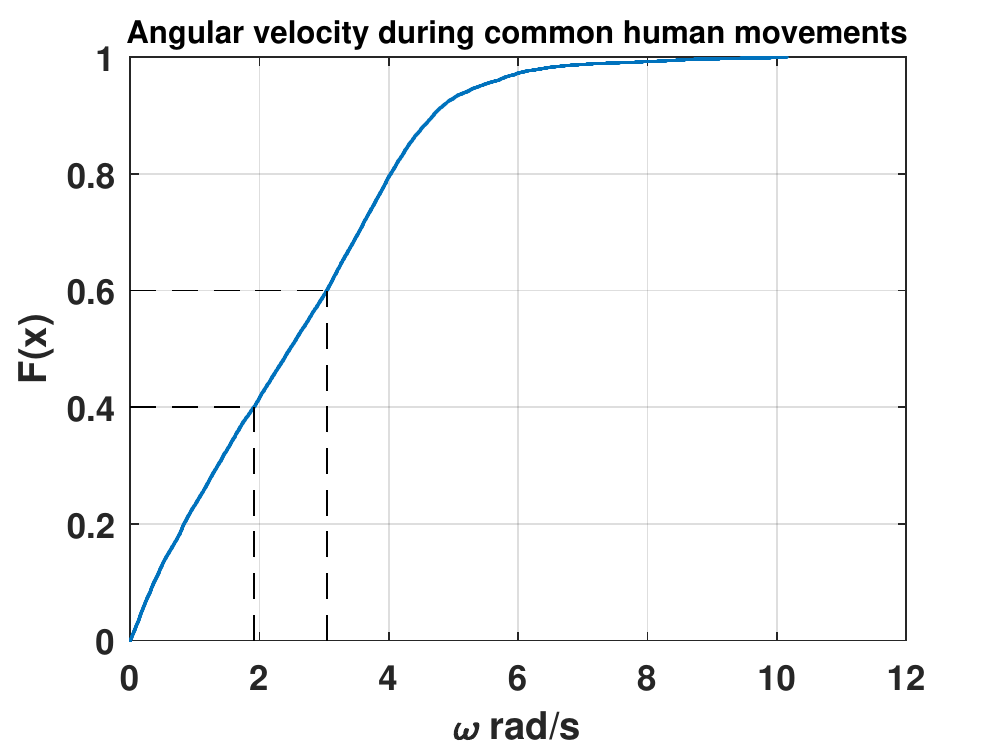}
    \caption{Gyroscope data during daily activities}
    \label{gyroscope}
\end{figure}

\textbf{Rotational Mobility:}
Rotational mobility disrupts beam alignment between the base station and mobile faster than translation mobility \cite{twc_beamsurfer,trackmac}. The alignment of the beams therefore lasts for a shorter duration, and
the mobile side receive beam can listen to a base station beam only for a shorter period. We experimentally investigated the expected large variation in the number of beams searched and the received signal strength on the aligned beam. 
 Before the experiments, we first observe angular velocity from the gyroscope on a commercial mobile device \cite{stmicroelectronics}. We logged data for one day. To name a few motion patterns during data collection are answering a phone call, walking with the mobile in hand, sitting on a chair, etc. Fig. \ref{gyroscope} shows the CDF of the angular velocity of the mobile.
We observe that angular velocity can reach up to 8 $rad/s$. We rotated phased array in our experiment with $40^{th}$ and $60^{th}$ percentile angular velocities from the observed data i.e., $90^\circ/s$ and $180^\circ/s$ and performed spatial scan during the mobility. 

\begin{figure}
     \begin{subfigure}[]
      \centering
         \includegraphics[width=.45\linewidth]{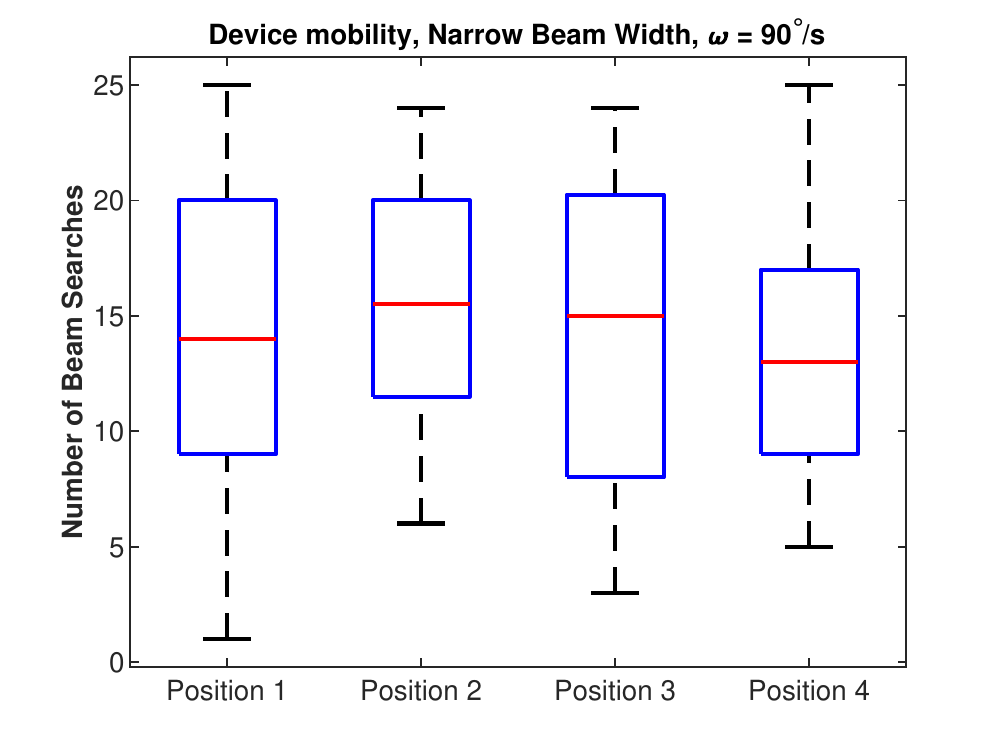}
    \end{subfigure}
           \hfill
     \begin{subfigure}[]
     \centering
      \includegraphics[width=.45\linewidth]{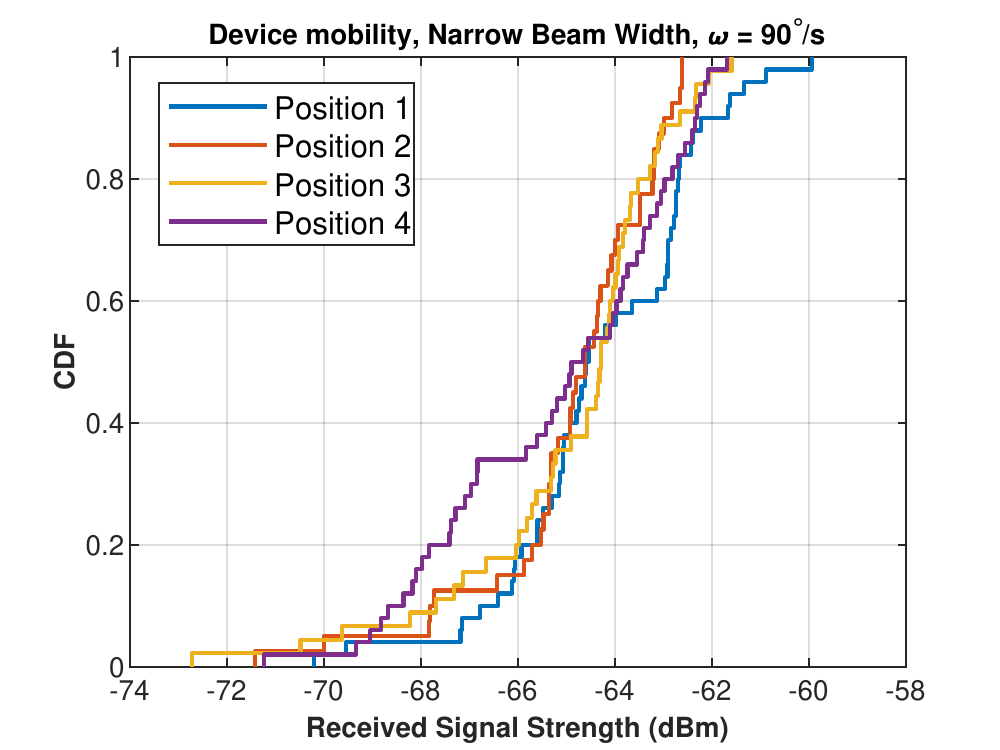}
     \end{subfigure}
     
     \begin{subfigure}[]
      \centering
         \includegraphics[width=.45\linewidth]{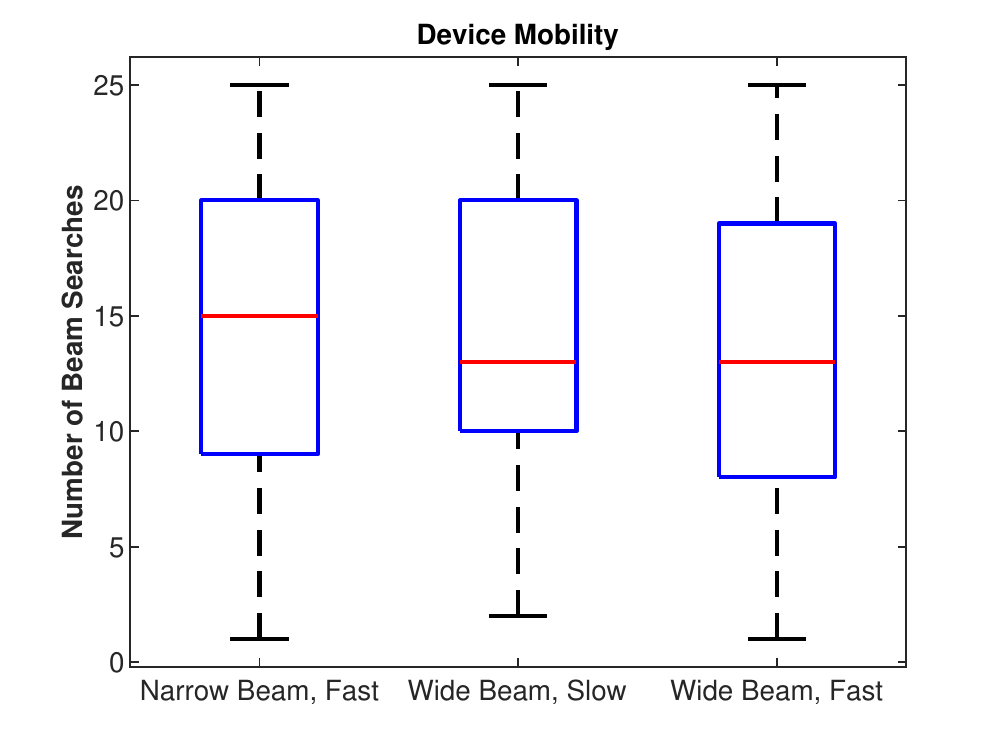}
           \end{subfigure}
         \hfill  
     \begin{subfigure}[]
     \centering
      \includegraphics[width=.45\linewidth]{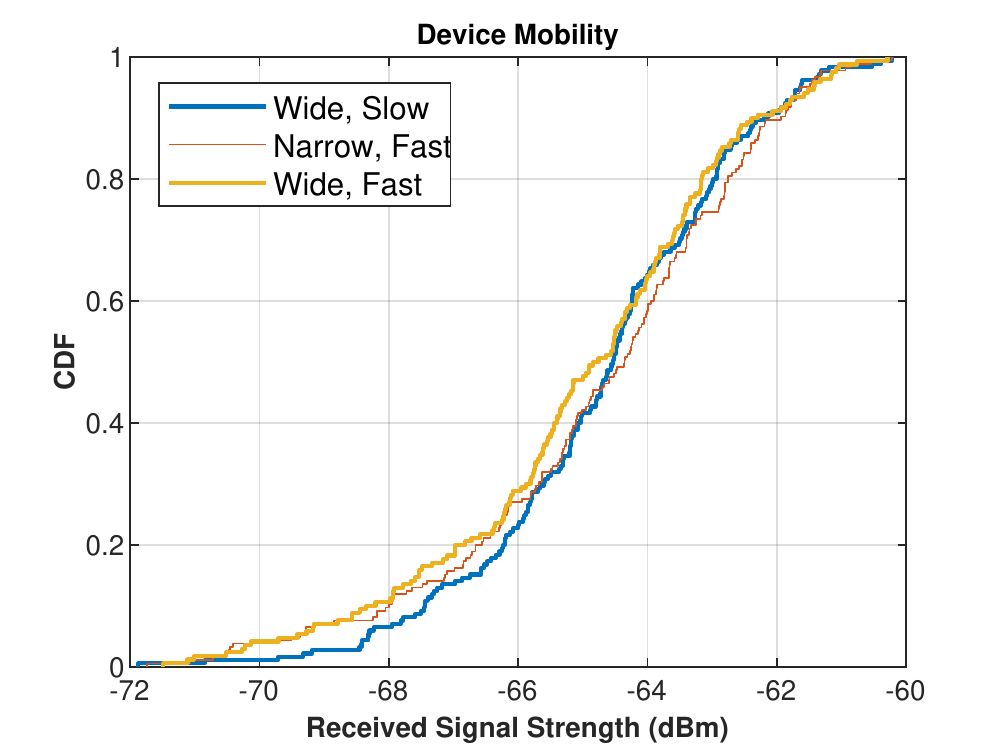}
     \end{subfigure}
     
    \caption{Impact of rotational mobility on  search}
    \label{rotationalmobility}
\end{figure}

In Fig. \ref{rotationalmobility}a, we present the number of beams searched for at four equidistant positions while a phased array rotates in a $120^\circ$ sector with angular velocities of $90^\circ/s$ and $180^\circ/s$.  As observed in Figs. \ref{lt_boresight_0}a, \ref{lt_boresight_60_120}a, \ref{rotationalmobility}c, and d,  the standard deviation of the number of beams searched during rotational mobility, 6, is higher than the corresponding number of 2 under the linear translational motion.

\textbf{Walk:} First we studied walking on a linear trajectory, next we looked at rotational motion, but human walk often has both rotational and translational components. So, we studied free walk, wherein a human walks casually, freely turns and changes direction of the motion.

We repeated the search experiments during such ``casual walks". We have also repeated the such experiments near two base stations to see if search is any faster under such deployments.

\begin{figure}
\begin{subfigure}[]
      \centering
         \includegraphics[width=.45\linewidth]{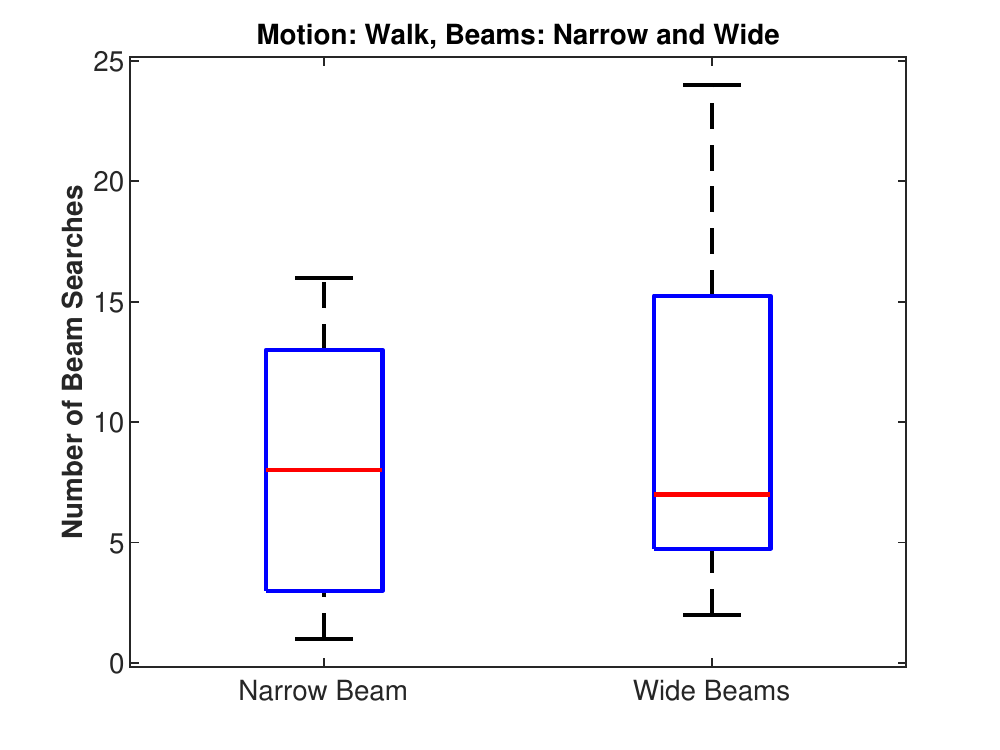}
    \end{subfigure}
           \hfill
     \begin{subfigure}[]
     \centering
      \includegraphics[width=.45\linewidth]{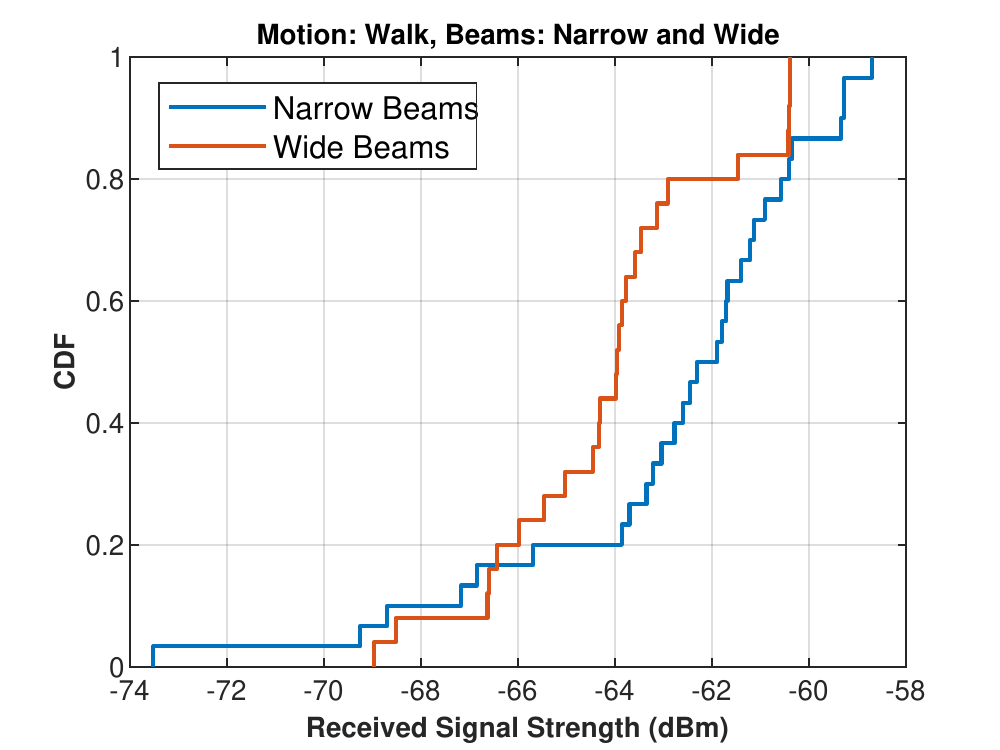}
\end{subfigure}
\caption{ Beam search while walking near one neighbor and two neighbor base stations a) number of beam searched b) received signal strength}
\label{freewalk}     
\end{figure}

\begin{figure}

 \begin{subfigure}[]
      \centering
         \includegraphics[width=.45\linewidth]{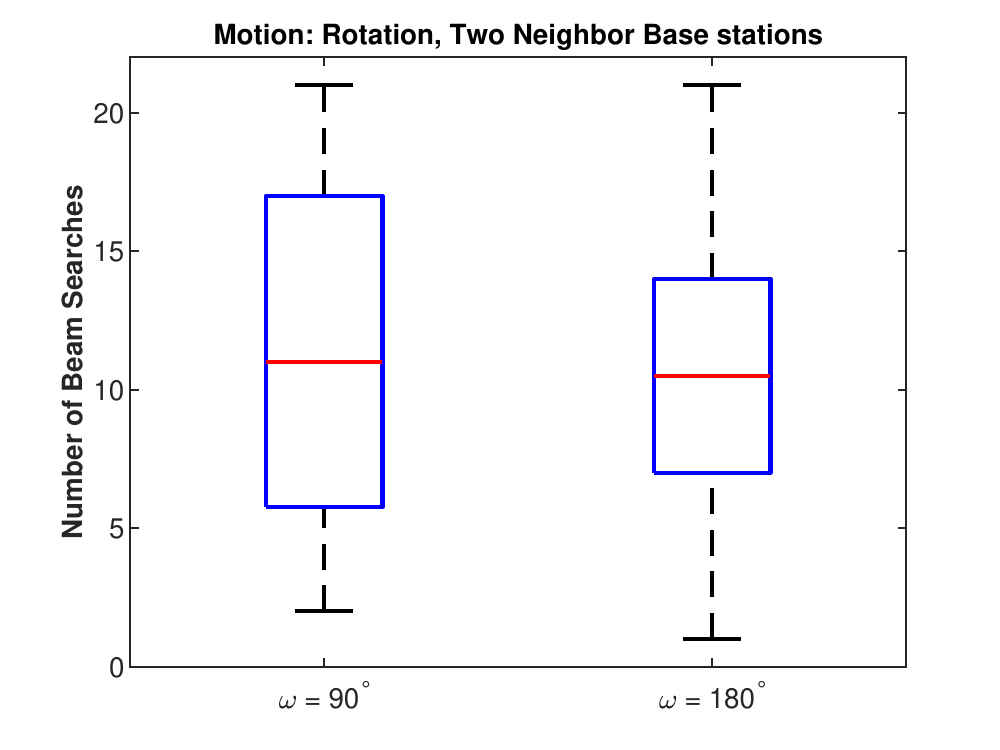}
    \end{subfigure}
           \hfill
     \begin{subfigure}[]
     \centering
      \includegraphics[width=.45\linewidth]{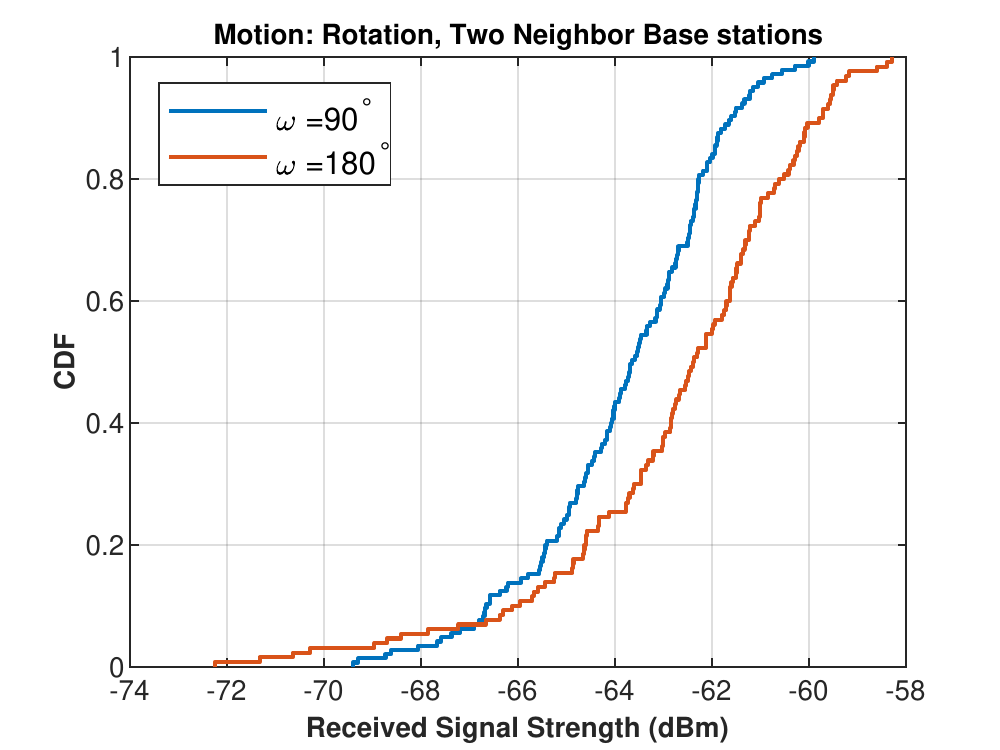}
     \end{subfigure}
    \caption{Beam search while rotating mobile near one neighbor and two neighbor base stations a) number of beam searched b) received signal strength }
    \label{rotate_2b}
\end{figure}

 Fig. \ref{freewalk} a $\&$ b present search overhead and received signal strength during walk. We observe slight decrease in search overhead when two beams from beams are visible to the mobile. mobile discovers a base station quickly. Similar observations can be made from Fig. \ref{rotate_2b}a where we repeated rotational mobility experiments near two base stations. Prior art \cite{Ish_blockage} also found dense mm-wave network deployments helpful.

\subsection{Tracking the Neighbor Base Station Beam}

After completing the spatial scan, it is ideal for mobile to not to repeat search, as it is both time consuming and energy intensive. To accomplish that mobile must keep track of the transmit beam of the found base station until handover is complete. Tracking involves the mobile adapting its receive beam to maintain alignment with the base station beam. Prior works have established that the alignment of base station and mobile-side beams 
degrades over time as the user moves \cite{beamsurfer,beamcoherencetime1,trackmac}.    
% While the user is moving, the earlier found beam needs adaptation. 
Unlike connected mode beam tracking, where the serving base station can help with beam adaptation,  the mobile does not get any help from the neighbor station. The mobile must solely and continuously track the discovered base station beam until the connection transfer is complete. 

% Works \cite{beamsurfer,beamcoherencetime1} investigated beam coherence time, the duration after which alignment between a beam pair gets disrupted with user mobility. As stated before in Section \ref{challenges}, beam adaptation methods that rely on the estimation of the angle of arrival/departure for beam tracking are erroneous at low SNR, a characteristic of cell edge. Therefore we need an in-band, minimal processing-based quick beam tracking mechanism.

We have conducted experimental investigations to study whether the highly aligned mobile-side receive beams follow any particular pattern as the user moves.
Several mobility patterns are studied, with the mobile's phased array's rotated with angular velocity $60^\circ/s$, $120^\circ/s$ and $240^\circ/s$ as well as natural walk.
For 100 experiment trials, we recorded the highly aligned mobile side receive beam directions. In each trail, there are 100 beam directions that provided highest receive signal strength at the mobile. These sequences represent the best aligned LoS directions in an experiment trial.

Our observation is that at any given point in time in the experiment, the best beam direction is an angular neighbor to past receive beam. For better visualization, suppose if we replace each element of the sequence with the angular distance from the previous element, sequence is a simple random walk. To further illustrate this observation we performed statistical tests \footnote{To test the randomness hypothesis, we performed runs-test on the sequences \cite{bradley_1968} with $95\%$ confidence interval and the null hypothesis being the sequence is random. } of randomness on the 100 sequences. The hypothesis tests failed to accept that sequences are generated from a deterministic source and hence might have random pattern. This implies that relying on the historical aligned beams to predict next beam is of no help. As sequence is a simple random walk, to track the base station beam it is sufficient for the mobile to check the signal strength on the angular neighbors of the current beam best beam direction.

\begin{comment}

\begin{table}[H]
\small 
    \caption{Runs Test \prk{This table does not help the reader. It requires way too much technicalities to understand. You need a different simple table or figure that is trivial to understand the bottom line. You can say as a side remark that you have conducted statistical texts, and they support the conclusions you are making.} \sg{removing table}}\label{runstest}
	\centering

		\begin{tabular}{ |P{4 cm}|P{1cm}|P{3cm}|}
\hline		
\hline
    {Motion } & {H} & {P-value}\\
 \hline\hline
 $\omega$ = 60 (deg/s) & 0  & .541 \\ 
 \hline
$\omega$ = 120 (deg/s) & 0 &.721\\ 
 \hline

$\omega$ = 240 (deg/s) & 0 & .128 \\ 
 \hline

Walk & 0  &.247\\ 
 \hline
\hline
\end{tabular}
\end{table}

\red{You don't need a whole section for this.
You can just write a brief paragraph at the end of the previous section.}

\end{comment}

Following our experimental observations, below is the summary of Beam Management of TERRA protocol

\textbf{Blockage Beam Management}: Contrary to the presumed requirement of the dense mm-wave base station deployment to mitigate blockage in outdoors, we found a simple alternative. By switching beam in the direction of ground reflection, mobile can maintain connectivity during temporary pedestrian blockage in outdoor environments. Received signal strength on the ground reflected beam direction is at least 6 dB less than LoS and is sufficient to maintain connectivity as well as low data rate control plane traffic.

\textbf{Handover Beam Management}: Search for a neighbor base station during user mobility is delay prone, consequently mobile needs a head start in search. To perform soft handover, mobile must have found a reserve base station and keeping track of its beam prior to switching decision.
To track the found beam i.e., to maintain an aligned receive beam with neighbor base station, the mobile  can only rely on the receive beam adaptation.
 The history of aligned receive beams cannot help mobile predict the next beam to employ to track the neighbor base station beam. Given the currently aligned receive beam, mobile can only statistically keep track of a neighbor base station beam. We show that by following the approach proposed in \cite{twc_beamsurfer}, that reduces the beam search space to the neighbors of the currently aligned receive beam, mobile can keep track of neighbor base station until handover is complete.
    
We present an in-band beam management protocol called TERRA in Section \ref{protocol_section} that manages beams in outdoor environments.
We present its efficacy both under temporary blockage and during handover in Section \ref{evaluation_section}

%% file: protocol.tex
\section{The TERRA Protocol for Transient Blockage and Handover}\label{protocol_section}
\begin{figure}[H]
    \centering
    \includegraphics[width=.95\linewidth]{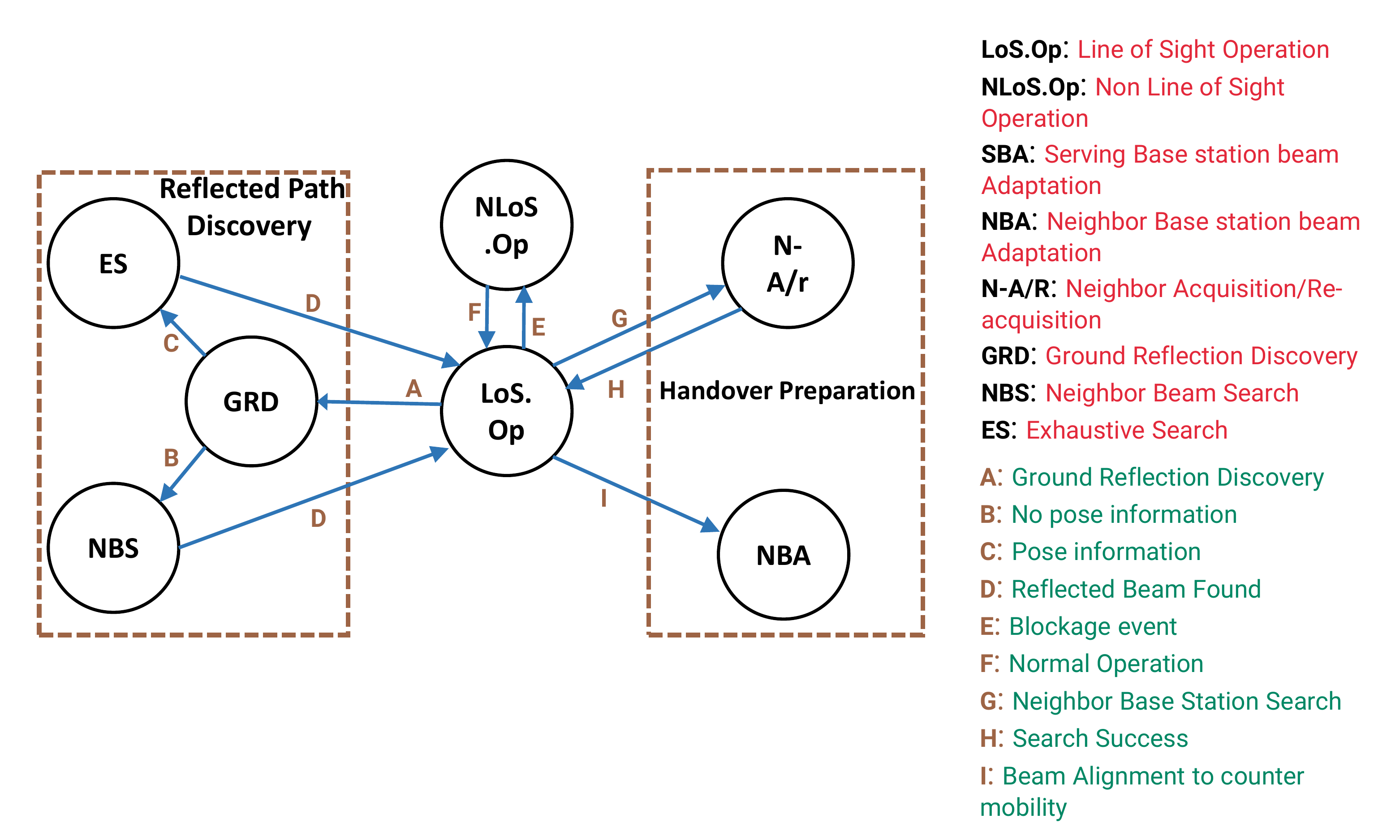}
    \caption{TERRA state machine}
    \label{protocol}
\end{figure}

{In this section we present the TERRA protocol that maximizes connectivity
in transient blockages as well as conducts soft handover over directional beams if needed.} Fig. \ref{protocol} presents the state transition diagram of the TERRA protocol.

A mm-wave mobile must continuously adapt LoS beam with the serving base station to counter user mobility. In LoS Operation State,  mobile adjusts its receive beam to maintain high degree of alignment with serving base station beam. Mobile may use beam alignment method proposed in \cite{twc_beamsurfer} to adjust LoS beam with serving base station to counter user motion. In addition to that, the mobile needs a backup NLoS beam to avoid disconnection during the temporary pedestrian blockages. TERRA protocol ensures this in Ground Reflected Beam Discovery (GRD) state wherein mobile identifies a ground reflected beam direction that we showed to have usable received signal strength. If pose of the mobile is available to the protocol, mobile moves to Neighbor Beam Search state to search zenith neighbors to current LoS beam and  discovers the ground reflections. When the knowledge of pose of the mobile is not available, TERRA protocol initiates search using all the receive beams in Exhaustive Search (ES) state to discover backup NLoS ground reflected beam. Ground reflected direction changes whenever mobile adapts its LoS direction. Therefore, TERRA must rediscover ground reflected direction too. Protocol erases the stored reflected direction in memory after LoS beam adaptation and visits GRD state to identify the ground reflected path. Protocol detects the transient blockage event when the received signal strength suddenly decreases by 15 dB as shown in Fig. \ref{blockage_eval}a, and in such an event TERRA employs ground reflection and keeps RSS within 6 dB of RSS on LoS beam.

 Unlike transient pedestrian blockage event where mobile can switch to LoS path after briefly operating on NLoS path, during a permanent blockage event, when the  LoS beam is occluded by a building, tree or any other immovable obstacle, mobile must switch base station. Blockage either transient or permanent is sudden, only difference is that, it is possible to switch to LoS beam after transient blockage.  In any case, if the mobile decides to switch to neighbor base station, TERRA provides mechanism to do so. TERRA protocol searches for a neighbor base station and keeps track of a neighbor base station to quickly perform soft-handover.
\begin{comment}
\prk{Then describe the stages of how handoff happens.
You will also need to say how a mobile finds out it is in a cell edge state.
This may require some additional steps in the State Diagram.
For permanent blockage induced handoff, will also need to add an explanation of how a mobile discovers that blockage could be permanent (example by a building), necessitating a handover. This will probably need a timeout.
}

\prk{As soon as the mobile detects deteriorating radio conditions, which we call a ``Cell Edge" operation, the mobile starts searching for neighbor base stations. Isn't this done always? A permanent blockage by a building could happen anytime,
and the mobile will need to be prepared to handoff right away.}

\sg{TERRA protocol helps a mobile to discover and keep track of the neighbor base station beam. The policy on whether or when to switch base station is left open. Goal of the TERRA protocol is to enable the mobile to efficiently manage beams in outdoors.} 
\end{comment}

To search for a neighbor base station, the mobile moves to the Neighbor Acquisition/Re-acquisition state (N-A/R). In this state, the mobile performs  spatial scan and discovers at least one neighbor base station beam. Mobile also identifies a receive beam, which we call the ``current receive beam of neighbor station" on which it can listen to the found base station beam.

The mobile needs to adapt its receive beams to counter user mobility during handover and monitor the  neighbor base station beam. 

Mobile employs the only viable choice, the receive beam adaptation,
to track neighbor base station beam, in the Neighbor base station Receive Beam Adaptation (N-RBA) state.  In this state, whenever the received signal strength of neighbor base station beam drops by 3 dB, mobile tests received signal strength on all the spatial neighbor beams to the current receive beam and chooses the beam improves the RSS. While the mobile maintains connectivity with serving base station and tracks the neighbor base station beam, handover maybe  initiated to a neighbor when the received signal strength exceeds hysteresis thresholds. This work focuses only on beam management during handover and there is extensive prior work on switching criteria for handover.

%% file: evaluation.tex
\section{evaluation}\label{evaluation_section}
We evaluate TERRA protocol under pedestrian blockages and its ability to track a neighbor base station beam. The experiments and evaluation although performed on our 60 GHz National Instruments transceiver \cite{instruments_2020_introduction} can be reproduced on any available mm-wave hardware. We provide a python notebook \cite{mikrotik} to control the beams of an off-the-shelf and cheaper mm-wave  hardware to help researchers reproduce the work. 

\subsection{Blockage Recovery}

\begin{figure}
\begin{subfigure}[]
      \centering
         \includegraphics[width=.45\linewidth]{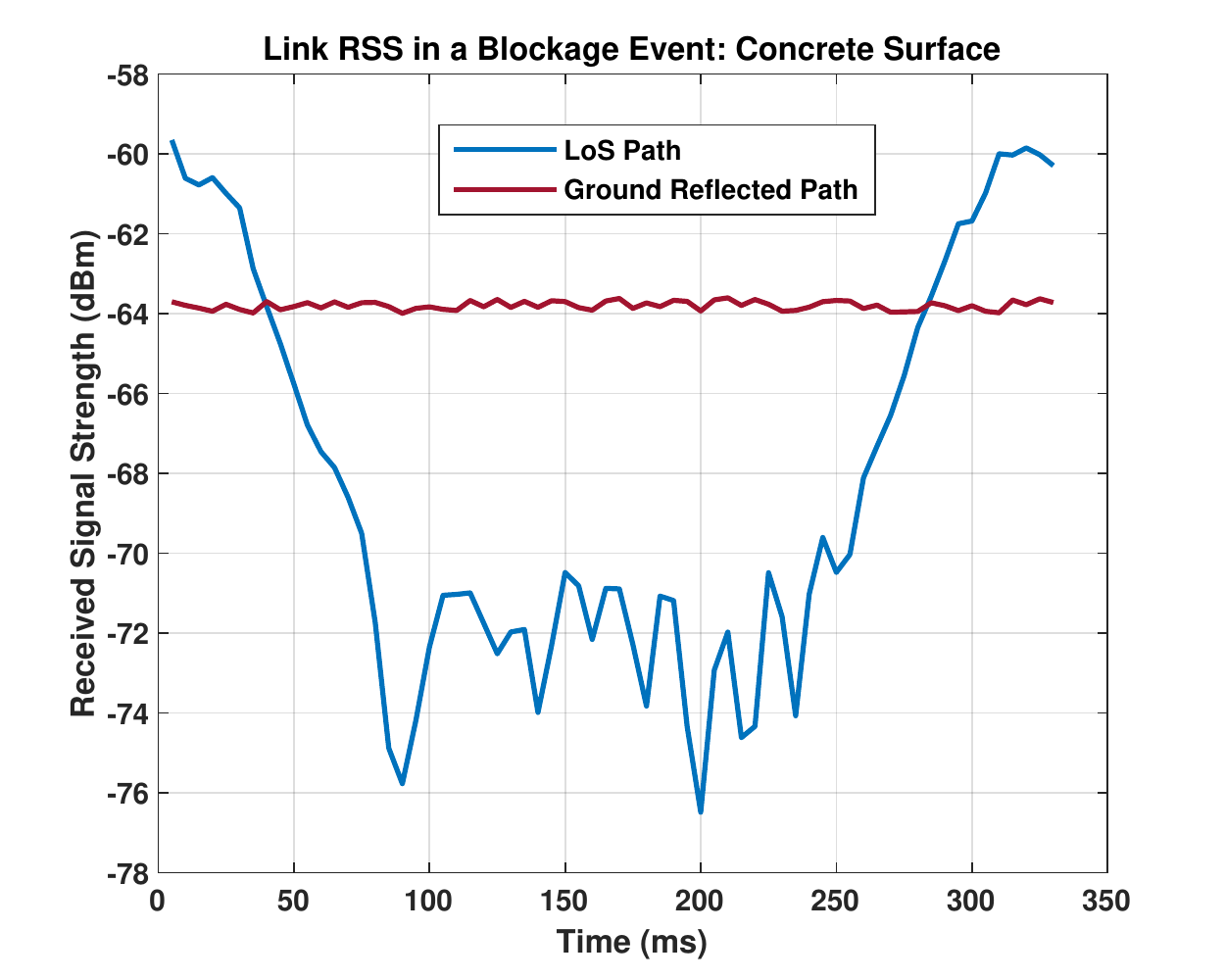}
    \end{subfigure}
           \hfill
     \begin{subfigure}[]
     \centering
      \includegraphics[width=.45\linewidth]{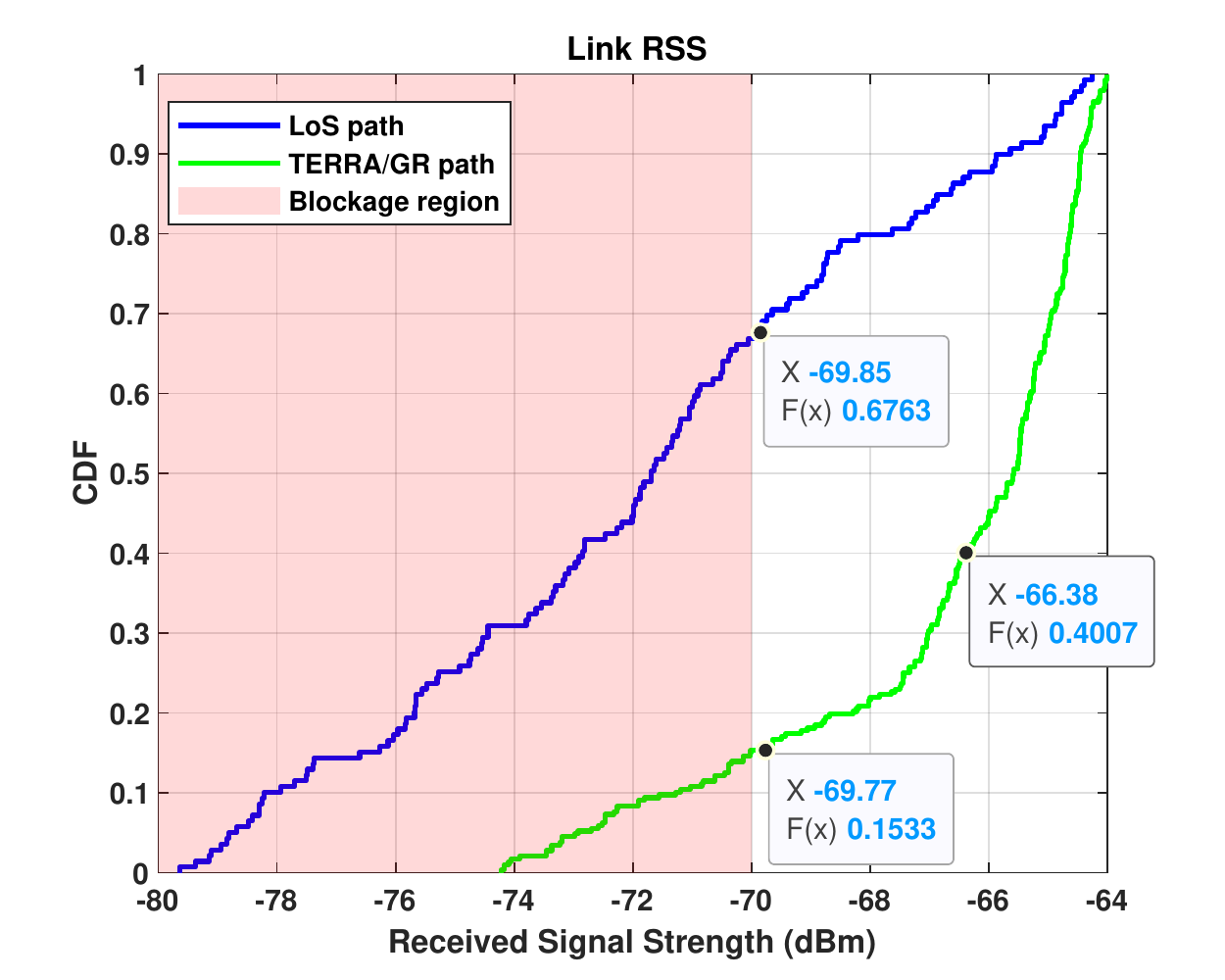}
     \end{subfigure}
    \caption{(a) RSS in a blockage event (b) Performance of TERRA during pedestrain blockages}
    \label{blockage_eval}
\end{figure}

Fig. \ref{blockage_eval}a shows the received signal strength of the LoS dropping below the noise floor of NI's 60 GHz receiver \cite{instruments_2020_introduction}, i.e., -70 dBm, during a pedestrian blockage that lasts for about 200 ms. Receiver experiences outage during this event as it cannot decode transmitted information. However, the ground reflected beam did not suffer outage. In this particular experiment trial on concrete surface, received signal strength is -64 dBm in ground reflected direction and  -60 dBm in LoS. We repeated our pedestrian blockage experiments at different locations with concrete and gravel surfaces.
% On the particular surface under test,
% whenever pose information is available, 

Fig. \ref{blockage_eval}b plots the CDFs obtained by employing TERRA's blockage recovery mechanism during 50 blockage events.
The colored region shows the
% the RSS in the LoS path 
outage region. The experiments show that TERRA employs beams that are
outside the outage region $84.5\%$ of time, and within 6 dB of normal operation $60\%$ of the time. When pose information of receiver is available, 
TERRA either finds ground reflected radiation in just two measurements, or else it searches all the available 25 beams till successful.

\subsection{Neighbor Base Station Beam Tracking}
\begin{figure}
\begin{subfigure}[]
      \centering
         \includegraphics[width=.45\linewidth]{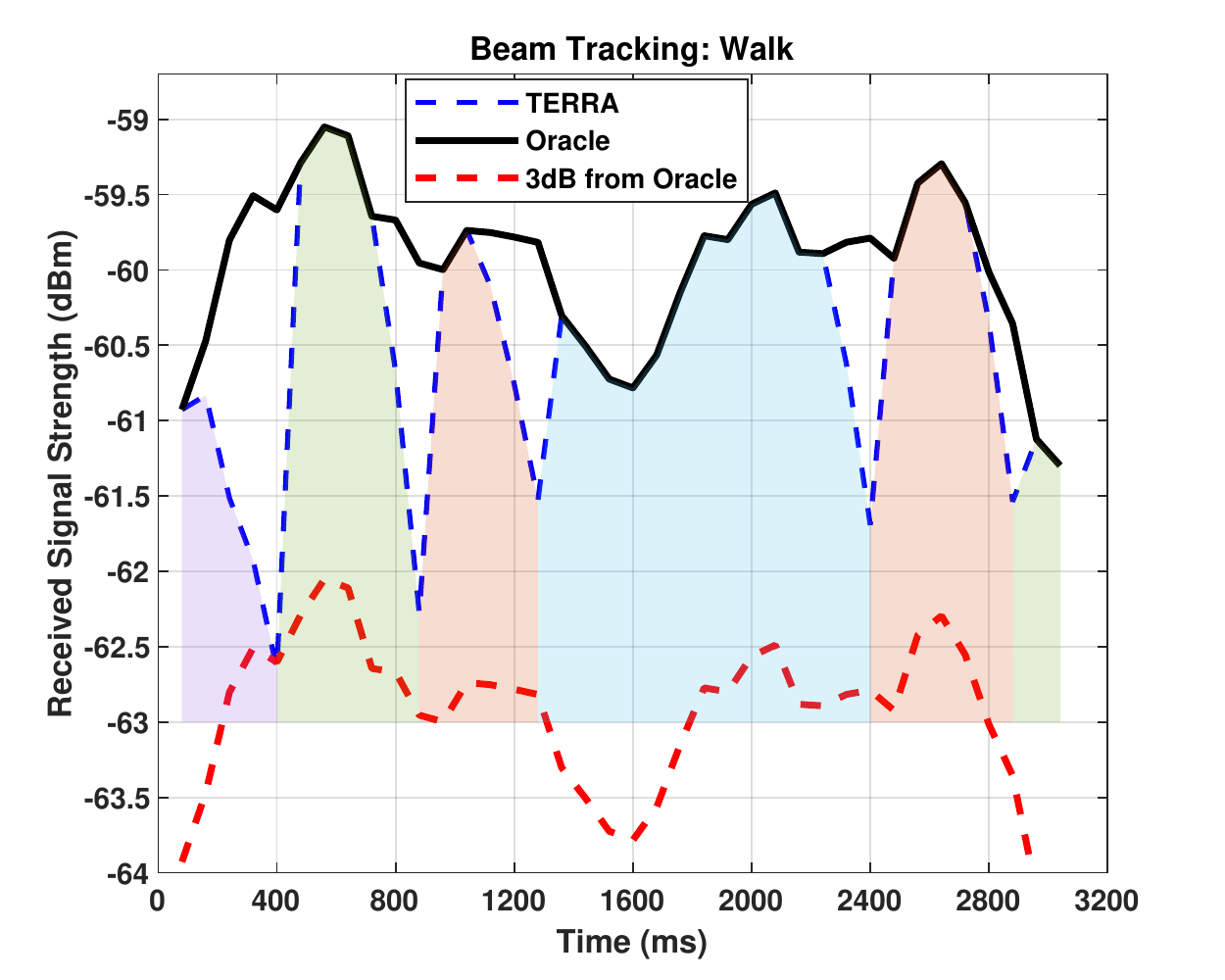}
    \end{subfigure}
           \hfill
     \begin{subfigure}[]
     \centering
      \includegraphics[width=.45\linewidth]{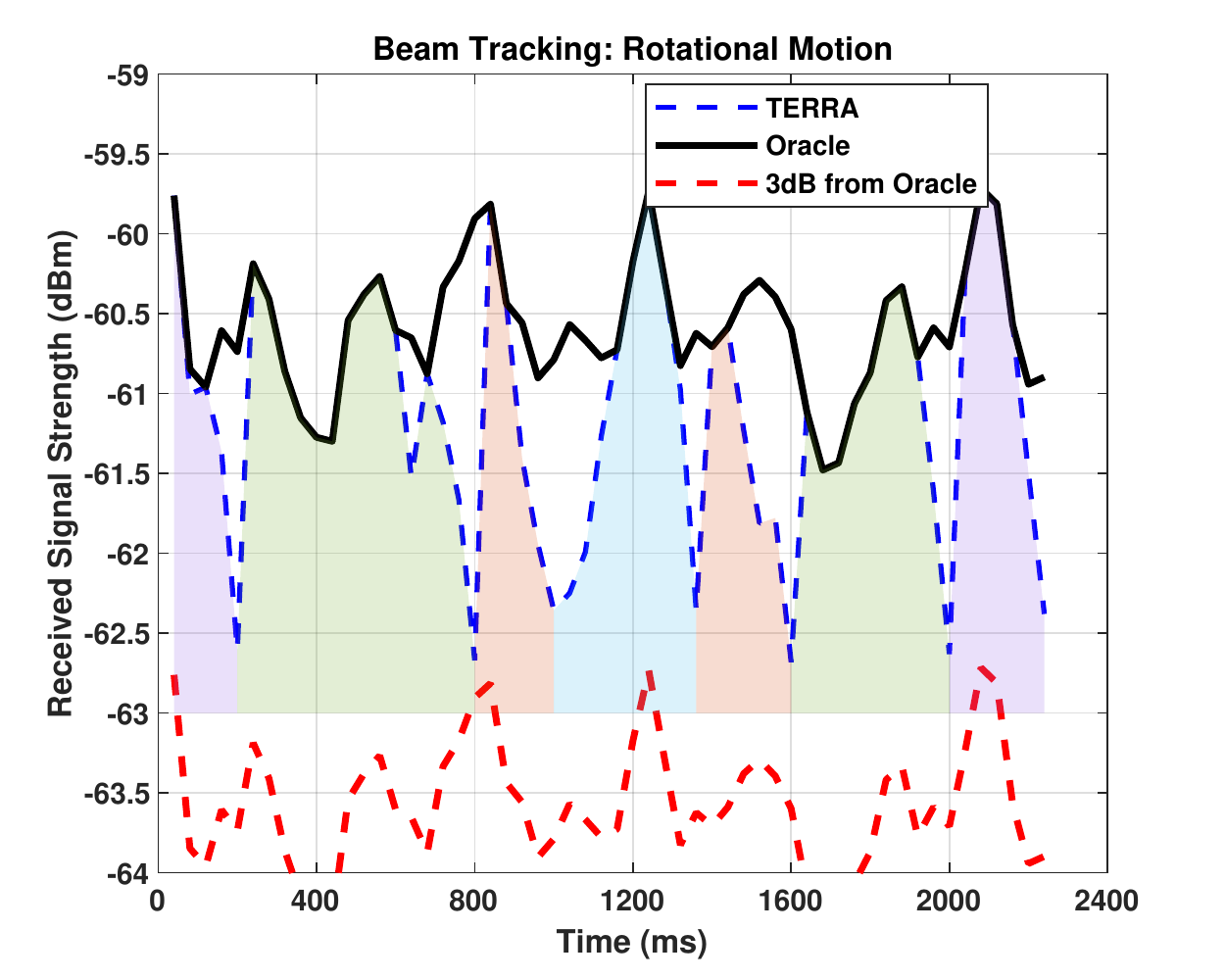}
      \end{subfigure}
\caption{ Beam tracking performance: a) Walking b) Rotation Motion $\omega$ = 120 deg/s }
\label{tracking_performance}     
\end{figure}

Tracking the neighbor base station beam is crucial for successful soft handover. TERRA
 actively tracks the neighbor base station. In this section, we present the experimental performance 
evaluation of TERRA. 
% We performed mobility experiments to show the beam tracking performance of TERRA. 
We conducted 50 trials of each of the mobility pattern mentioned in Table \ref{rssdeviation2}. TERRA starts with the receive beam found after the initial search and monitors the received signal strength on that beam. 
Once the received signal strength goes down by 3 dB due to mobility, TERRA switches its receive beam direction to an adjacent  beam that improves received signal strength. It then continues to monitor received signal strength using the new beam. Fig. \ref{tracking_performance}a shows received the signal strength at the mobile in one of the trials of our experiments where the user is walking near the edge of the cell. 
The different beam directions are colored differently in Fig. \ref{tracking_performance}a and b,
thus showing the beam switching resulting from the TERRA protocol whose performance is indicated in dashed line. 

We also present in Fig. \ref{tracking_performance}a and b the performance of an ``Oracle"
that has omniscient knowledge of the received signal strength of all the codewords and chooses the maximum.
To measure Oracle received signal strength, we swept all the receive beams in the beam codebook and took the maximum.
A parallel curve representing a 3dB loss over Oracle is also shown  in Figs. \ref{tracking_performance}a,b.
As can be seen, for 3 seconds of the in the Walk experiment, and  2.4 seconds in the Rotational Motion experiment, the received signal strength observed using TERRA is within 3 dB of the Oracle signal strength. Also, we can see that each receive beam of the TERRA takes advantage of the entire main lobe before switching to the next. Although Oracle signal strength is above TERRA's, it is achievable only if the mobile exhaustively searches all its beams every time. 
% To evaluate how Silent Tracker is performing with respect to the best beam that it could have chosen if it had omniscient knowledge of the RSS of all codewords, The oracle received signal strength is the maximum received signal strength observed in our setup. 

\begin{table}[H]
\small 
   \caption{deviation of the TERRA's received signal strength from oracle solution.}\label{rssdeviation2}
       \centering

		\begin{tabular}{ |P{8.5 cm}|P{6.5cm}|}
\hline		
\hline
    {Motion } & {Root Mean Square Loss (in dB) } \\
 \hline\hline
 Rotational motion at $\omega$ = 60 (deg/s)  & .4247    \\ 
 \hline
Rotational motion at $\omega$= 120 (deg/s)  & .71 \\ 
 \hline
Walk: User holding phased array walks near base station & .8343  \\ 
 \hline
\hline
\end{tabular}
\end{table}
% In Fig. \ref{tracking_performance}b, we show the performance of TERRA during a rotational mobility experiment. The user with a phased array in her hand rotates the phased array, and TERRA adapts the receive beam to maintain good received signal strength. 
% Oracle received signal strength is maximum after measuring using all the beams in our codebook. 
% Similar to Fig. \ref{tracking_performance}a, TERRA ensures the received signal strength is within 3 dB from the oracle signal strength. 

 Table \ref{rssdeviation2} tabulates the deviation of TERRA's received signal strength from the Oracle in all mobility experiments. TERRA can be seen to maintain received signal strength within 0.5 dB during slow rotational motion. In a faster rotational experiment with angular velocity 120  $deg/s$, it is 0.71 dB. For pedestrian mobility where a human walks near the neighbor base station holding phased array in hand, it is 0.83 dB. Overall, in all the mobility patterns that we have studied, tracking performance is very close to Oracle, within 1 dB. 

\subsection{Simulation Study of Beam Tracking:}
Using computer simulations, we compare TERRA's beam tracking complexity with state-of-the-art beam alignment schemes {in terms of the} number of measurements necessary to identify a receive beam.

The simulation study uses a receiver with a uniform planar array with 32X32 elements, operating at 28 GHz, and is 10 m away from the transmitter. The total number of possible receive beams is 1024. Exhaustive search needs 1024 measurements to identify a receive beam that is highly aligned  with a transmitter beam. Receive beam adaptation is necessary when the current receive beam is no longer adequate.

The Hierarchial Beam Alignment (HBA) \cite{HBA} method first measures with an Omni-directional beam, and narrows down the beamwidth after each measurement. HBA \cite{HBA} employs correlated bandit learning to narrow down the beam. In our study, HBA \cite{HBA} needed 63 measurements to find an aligned receive beam. Given the sparse nature of the mm-wave channel in the spatial domain, compressive-sensing based approaches have been studied in literature to quickly identify aligned beams; fast beam alignment with low-resolution phase shifters (FALP) \cite{FALP} uses a variation of the compressive sensing method.

FALP \cite{FALP} requires 70 measurements to align the receive beam. Another approach called Agile Link \cite{agilelink} uses carefully designed beam patterns that have multiple lobes to receive the signal from multiple directions. This design gives Agile Link \cite{agilelink} the ability to search in multiple directions in one measurement. 
% Still, Agile Link needs several measurements \red{How many?}  to detect transmit beam direction.
In our simulation study Agile Link \cite{agilelink} took 110 measurements to find a receive beam that aligns with the transmit beam. In contrast, TERRA searches only the neighbors to the previously aligned beam, and
yet finds a receive beam that provides received signal strength within 3 dB of the Oracle beam. While Oracle needs an exhaustive search, TERRA takes a maximum of 8 searches to find an aligned receive beam.

\begin{table}[H]
    \centering
   \begin{tabular}{ | l | c |  }
\hline
\hline
State of the art & Maximum number of measurements  \\ \hline \hline
Terra & 8  \\  
 HBA & 63 \\
 FALP & 70 \\
 Agile Link & 110 \\
 Exhaustive Search & 1024 \\
 \hline
\end{tabular}
\caption{Tracking overhead}
\end{table}

%% file: related_work.tex
\section{Related work}
When the LoS beam of mobile is blocked for prolonged duration or when the user moves to the edge of the serving cell, mobile switches base stations. 
Broadly, beam management methods proposed so far to handover rely on location of the user;  Reinforcement Learning; Machine Learning.  Much of the prior work used computer simulations and a thorough experimental work addressing all the challenges in handling beams during a handover is missing, and TERRA protocol fills the void.

Significant number of works in the literature proposed either dense deployment of base stations or co-ordinated multiple point access (CoMP) to address outdoor pedestrian blockage. Recently, using reflective surfaces near the base stations in mm-wave deployments have been under investigation to overcome mm-wave link impairments. Mezavilla et al. have studied the use of such surfaces \cite{RISassisted}.   To the best of our knowledge, we first evaluated ground reflections to address temporary blockage in outdoor environments.

\textbf{User location:} Junshen et al. have proposed a method \cite{Beamsearch1} to reduce the numbers of beams to search to discover the target base station for handover.  Along with the geometry model of  environment, the approach requires co-ordination among current and target base station. Parada et al. have proposed a method \cite{directionofpass} using  user's direction of motion to predict the target base station for handover. Using a multitude of access points, Palacio et al. have developed a user localization algorithm \cite{location} and proposed location aware methods for beam adaptation and handover decisions. TERRA doesn't rely on user location, instead it performs a search to discover a neighbor base station and keeps track of the found beam with little overhead as shown in Section \ref{evaluation_section}.

\textbf{Reinforcement Learning:}
To learn a good beam management policy i.e., which beam and the base station to switch to, 
reinforcement learning based methods require precise and high fidelity model of the environment. A learning agent using any of the popular reinforcement learning algorithms interacts with environment model and learns a beam management policy. Additionally, one must do fine reward engineering that helps agent evaluate the actions for every state of the mobile in a given environment. State of the mobile may include signal strength, signal-to-noise ratio, user location, and speed of the mobile.

Zang et al. have employed model-based reinforcement learning \cite{learning1} to learn a beam management policy for handover. The model uses the mobile's location, velocity, and connection state as state information, and uses a Gauss-Markov mobility model for the transition kernel. Sun et al have explored multi-arm bandit approaches \cite{bandit}. 
Adding location and direction of motion of the user as state information, Sun et al. also applied contextual bandits \cite{contextual_bandits1}. Other works \cite{Reinformentlearning2,Reinformentlearning3,location_velocity,MDP1} have applied various reinforcement learning algorithms to make handover decisions with state information being the location and trajectory of users. However, through our experiments, we found that given the current best aligned beam, the next best beam is its angular neighbor. Also, in a random walk, the learned policy is stochastic and suggests Reinforcement Learning agent to try the neighbor beams which is not different from TERRA protocol.  

Another shortcoming Reinforcement Learning approaches is that it is not clear how a learned policy in an environment may perform in other environments and developing such algorithms, called meta learning, is still an active area of research in reinforcement learning. To keep the protocol design simple, TERRA switches to first found base station.  

\textbf{Machine Learning:}
Authors applied \cite{LSTM} a popular sequence prediction method in machine learning literature, Long-Short Term Memory (LSTM), to predict the mm-wave link quality. To that end authors simulated a base station deployment in 200x200 $m^2$ area. Although not accurate, LSTM approached performed better than moving average in predicting blockage events. TERRA protocol can leverage link prediction methods and employ ground reflected path.  The challenge with data driven machine learning methods is that they may or may not  perform well outside the trained environment.   Kaya et al. utilized ray tracing to construct radio environment for a video feed from a traffic intersection in a major city in the USA,  and \cite{LSTM1} applied LSTM to predict beams to aid pedestrian and vehicular mobility. LSTM predicts top-2 with beams that gives 96 $\% $ prediction accuracy. From the plots in work, we found that algorithms predicts neighbor beam to past best aligned beam. TERRA achieves this prediction accuracy without the need of history of past beams.

\textbf{Conclusion:} In this work, we propose TERRA protocol that effectively manages mobile side beam in outdoors environments. TERRA ensures an outdoor mobile act quickly in a transient blockage event and tracks a neighbor base station to perform soft-handover. We present both experimental and simulated evaluation to show efficacy of the protocol. While TERRA helps mobile avoid outage in outdoors, still careful base deployment is a must to address crowded environments. In future, we plan to study optimal network deployment that maximizes visibility of ground reflections to the mobiles.